\documentclass[lettersize,journal]{IEEEtran}
\usepackage{amsmath,amsfonts}
\usepackage{algorithmic}
\usepackage{algorithm}
\usepackage{array}
\usepackage[caption=false,font=normalsize,labelfont=sf,textfont=sf]{subfig}
\usepackage{textcomp}
\usepackage{stfloats}
\usepackage{url}
\usepackage{verbatim}
\usepackage{graphicx}
\usepackage{cite}
\usepackage{stfloats}

\begin{document}

\title{\textit{MANOJAVAM}: A Scalable, Unified FPGA Accelerator for Matrix Multiplication and Singular Value Decomposition in Principal Component Analysis}

\author{Srivaths~Ramasubramanian (Student Member, IEEE), Anjali~Devarajan (Student Member, IEEE), Kousthub~P~Kaivar (Student Member, IEEE), Vibha~Shrestta (Student Member, IEEE), Shashank~D (Student Member, IEEE), Sowmyarani C.N (Member, IEEE), Govinda~Raju~M and K.S Geetha (Senior Member, IEEE)
\thanks{This work was conducted at RV College of Engineering, Bangalore, India.}%
\thanks{Srivaths Ramasubramanian, Anjali Devarajan, Vibha Shrestta, Kousthub P Kaivar, Shashank D, Sowmyarani C.~N, Govinda Raju M and K.S Geetha are with RV College of Engineering, Bangalore, India (e-mail: srivathsr.ec21@rvce.edu.in; anjalid.ec21@rvce.edu.in; vibhashrestta.ec21@rvce.edu.in; kousthubp.ec21@rvce.edu.in; shashankd.ec21@rvce.edu.in; sowmyaranicn@rvce.edu.in; govindarajum@rvce.edu.in; viceprincipal@rvce.edu.in).}%
}

\maketitle

\begin{abstract}
Principal Component Analysis (PCA) is widely used for dimensionality reduction in hyperspectral imaging, genomics, and neurosciences. However, it suffers from computational bottlenecks in matrix multiplication and singular value decomposition (SVD). Prior PCA hardware accelerators either target only one of these stages, rely on High Level Synthesis (HLS) that limits microarchitectural optimizations or use fixed point datapaths with limited dataset scalability. There is a need for a unified PCA accelerator that is suitable for datasets of any input dimension.
Hence, the proposed work presents MANOJAVAM, a scalable PCA accelerator fabric, unifying matrix multiplication and SVD in a single architecture. MANOJAVAM(T,S) comprises an S number of TxT TPU-style systolic arrays employing block streaming for high-throughput matrix multiplication. It further integrates a highly parallel Jacobian unit implementing the Jacobi method for SVD with pipelined CORDIC based rotations. A two tier cache hierarchy and mode-aware memory policies adapts to the distinct memory access patterns of covariance matrix and rotation computation. 
For demonstration, MANOJAVAM(4,8) is realized on a Xilinx Artix-7 FPGA, achieving a frequency of 200 MHz at 1.271W. MANOJAVAM(16,32) is realized on Xilinx Virtex-Ultrascale+ FPGA, achieving a frequency of 434 MHz at 16.957W. Benchmarking on real-world datasets reveals that MANOJAVAM(16,32) achieves up to a 22.75x speedup in SVD latency and a 42.14x reduction in total energy consumption compared to a high-performance NVIDIA A6000 GPU. The architecture offers a unified, scalable, and energy-efficient platform for large-scale data analytics in both high-performance and edge-computing environments.
\end{abstract}


\section{Introduction}
\IEEEPARstart{T}{he} end of Dennard scaling\cite{introduction_1} and the slowing of Moore's law\cite{introduction_2} have made it difficult to meet the growing demands of modern day computational workloads. Modern day CPUs, though extremely fast in operation, are infeasible solutions to accelerate compute-intensive tasks due to their sequential nature. A major challenge in general purpose processors is that they are power hungry, as they offer support for multiple operations, have excessive precision support and a higher frequency of operations \cite{introduction_4}. On the other hand, acceleration on Graphic Processing Units (GPUs) is largely constrained by very high power consumption and memory bottlenecks\cite{introduction_7}, thus limiting their full utilization of compute cores\cite{introduction_6}. Thus, there is a need to accelerate such workloads with domain specific architectures\cite{introduction_3}.

As a result, hardware acceleration has emerged as a promising solution to speed-up massively parallel tasks and extract very high performance\cite{introduction_5, introduction_6}. Field Programmable Gate Arrays (FPGAs) have emerged as a primary platform for such tasks, offering gate-level reconfigurability and a high degree of parallelism. This enables rapid prototyping and the ability to implement highly optimized, domain-specific datapaths that are not constrained by the fixed instruction sets of general-purpose processors. Consequently, workloads accelerated on FPGAs show superior performance and energy efficiency compared to their general-purpose counterparts.

Principal Component Analysis (PCA), is an important data mining technique used in several real world applications, such as image compression\cite{pca_applications_1}, image classification\cite{pca_applications_2}, facial detection\cite{pca_applications_3}, neurosciences\cite{pca_applications_4} and genomics\cite{pca_applications_5, pca_applications_6}.  PCA reduces the dimensionality of an input dataset by mapping it to a lower dimensional subspace. It involves the transformation of the dataset’s original features into a set of Principal Components, that maximally contain the variance of the input dataset while reducing the number of attributes\cite{pca_review_1, pca_review_2, pca_review_3}. The algorithm suffers from two bottlenecks: the computation of the Covariance matrix and Singular Value Decomposition (SVD). While software-based PCA implementations suffice for offline data analysis, they introduce non-deterministic latencies that are prohibitive for high-speed edge applications, such as autonomous drone navigation and real-time medical imaging. Existing general-purpose accelerators, such as mobile-GPUs, offer high peak throughput but often suffer from significant power overhead and memory bottlenecks during the iterative stages of eigendecomposition. 

Numerous attempts have been made to accelerate Principal Component Analysis\cite{pca_architecture-1, pca_architecture-2, pca_architecture-3, pca_architecture-4, pca_architecture-5, pca_architecture-6, pca_architecture-7, pca_architecture-8, pca_architecture-9, pca_architecture-10, pca_architecture-11}, employing different methods to expedite the computation of covariance matrix and singular value decomposition. Prior work however, have suffered high resource consumption and operate on fixed input matrix dimensions. In this work, we present a parametric architecture where accelerator performance can be tuned by adjusting parallelism ($S$) and tile size ($T$). While power dissipation naturally scales with these parameters, our baseline configuration of $T=4$ and $S=8$ achieves an low power profile of 1.271W on an Artix-7 FPGA, representing the lowest power consumption among the state of the art.

The proposed work presents \textit{MANOJAVAM}, a scalable and unified accelerator that integrates both matrix multiplication and SVD computation within a single architecture. The contributions of this work are as follows:
\begin{itemize}
    \item Unified Datapath: An accelerator that reuses its $T \times T$ systolic array cores for covariance matrix computation and Jacobi rotations, reducing resource utilization on FPGAs
    \item Integrated Jacobi Unit with DLE: The Jacobi unit features a Data Lookup Engine (DLE) which directly interfaces with the outputs of the covariance matrix to obtain the pivot elements in a single scan once the accumulation is complete. This removes the need for round trips to BRAM to extract these elements. The DLE also implements tile-aware filtering to ensure it looks up at valid output tiles in the search of the pivot elements.
   \item Numerical Robustness and Deterministic Convergence: A systematic Frobenius-norm-based evaluation across diverse datasets to establish a deterministic 50-sweep iteration schedule. This provides a significant safety margin for ill-conditioned datasets while eliminating the hardware overhead of complex on-chip convergence monitoring logic.
    \item Mode-Aware Memory Subsystem: A specialized two-tier cache hierarchy featuring two different cache write-miss policies to accommodate for distinct data patterns observed in covariance matrix computations and Jacobi rotations. 
    \item Multi-Platform FPGA Validation and Benchmarking: A comprehensive evaluation across the Artix-7 and the Virtex UltraScale+ FPGAs. Compared to a workstation-grade NVIDIA A6000 GPU, MANOJAVAM achieves up to a 28.2$\times$ speedup and over six orders of magnitude ($1.3 \times 10^6\times$) higher energy efficiency, while eliminating the non-deterministic software orchestration latencies native to GPU-based acceleration.
\end{itemize}
\section{Literature Survey}
Principal Component Analysis (PCA) remains a computational bottleneck in real-time and resource-constrained environments due to the high arithmetic and memory demands of covariance computation and singular value decomposition (SVD). A wide body of research has focused on accelerating portions of this pipeline using FPGAs, yet most designs treat the covariance and decomposition stages in isolation. This section traces the evolution of PCA hardware acceleration by progressively analyzing the architectural approaches to (A) covariance matrix computation, (B) SVD and eigendecomposition, and (C) unified designs that attempt to integrate both stages. We also survey (D) algorithmic strategies that impact hardware parallelism and convergence, leading to (E) the motivation for \textit{MANOJAVAM}.

\subsection{Covariance Matrix Acceleration}
The computation of the covariance matrix $C=X^{T}X$, has drawn early attention in FPGA-based PCA accelerators. These designs often adopt block or vectorized matrix multiplication schemes to improve throughput and memory reuse.

For instance, Korat et al. \cite{pca_architecture-1} presented an FPGA-based PCA pipeline that performs both learning and inference, using vector-matrix multiplication and QR decomposition for eigenspace extraction. While the design shows promise for small-scale matrices, its fixed-size datapaths and limited reuse strategy hinder scalability. Similar concerns are found in Mansoori et al. \cite{pca_architecture-3}, who developed an HLS-driven architecture with dedicated Covariance and SVD units for hyperspectral imaging; while modular, its resource use grows steeply with dimensionality, and the high-level synthesis approach offers limited microarchitectural flexibility.

A reconfigurable PCA accelerator was proposed by Shahrouzi et al. \cite{pca_architecture-5}, which was capable of dynamically switching between matrix multiplication and QR stages. This dynamic reconfiguration was carried out by programming the FPGA with pre-compiled bitstream files for matrix multiplication and eigendecomposition. However, the need to reload bitstreams introduced significant reconfiguration overhead, resulting in performance degradation. Fernandez et al. \cite{pca_architecture-2} sidestepped this issue by offloading matrix multiplication to a host processor, integrating only the eigendecomposition in hardware. While this reduces hardware complexity, it prevents full integration of compute stages. Consequently, intermediate results must be transferred between configurations, leading to I/O overhead and reduced overall performance. Das et al. \cite{pca_architecture-4}, on the other hand, applied PCA in FPGA-based network intrusion detection, where matrix multiplication was used to compute projections for dimensionality reduction. While their system demonstrates application-level effectiveness, it lacks on-chip decomposition logic and is tailored for fixed data characteristics, limiting its generality across domains.

\subsection{SVD and Eigendecomposition Acceleration}
Several works focus exclusively on accelerating the decomposition stage, often assuming a precomputed covariance matrix. These designs emphasize numerical stability, convergence speed, and hardware efficiency.

Ma and Wang \cite{svd_architecture-1} leveraged the Hestenes-Jacobi method using fixed-point arithmetic and CORDIC engines, demonstrating fast convergence for matrices under 128×128. Yet the design’s reliance on external tools and point-wise rotation logic hampers scalability. Chen et al. \cite{svd_architecture-3} tackled the decomposition of dynamically varying matrices in wireless communication by designing a reconfigurable SVD ASIC for MIMO-OFDM, integrating orthogonal reconstruction units. However, its serial rotation pipeline imposes latency constraints at scale.

Wang and Zambreno \cite{svd_architecture-6} pursued a general-purpose FPGA SVD engine using IEEE-754 floating-point units and Hestenes-Jacobi iterations. While it supports large matrices (up to 1024×1024), it lacks a matrix multiplication engine and relies on Xilinx IP blocks, making it less portable and energy efficient.

\subsection{Unified Architectures for Covariance + SVD}
Despite the natural interdependence of the covariance and decomposition stages in PCA, few hardware accelerators unify both phases within a single architecture. The integration of matrix multiplication, decomposition, and projection within shared resources introduces complex synchronization, memory pressure, and latency management challenges.

Among early efforts, Korat et al. \cite{pca_architecture-1} provided an end-to-end FPGA design for small matrices, but with minimal parallelism or scalability. Mansoori et al. \cite{pca_architecture-3} connected Covariance and SVD units in a modular HLS design but did not optimize data movement or exploit systolic structures. Shahrouzi et al. \cite{pca_architecture-5} enabled time-sharing of logic between stages, but this led to pipeline stalls and suboptimal throughput.

More structured integration appears in Zhang et al. \cite{svd_architecture-2}, who used Brent-Luk systolic arrays and Jacobi rotations to construct a tightly coupled SVD engine on FPGA. While effective for 8×8 matrices, their design peaked at 98\% resource utilization and could not scale to higher dimensions. Athi et al. \cite{svd_architecture-5} proposed a fast-converging parallel Jacobi method that accelerates convergence through strategic rotation ordering, yet the lack of a dedicated matrix multiplication core restricts its use in full PCA pipelines.

\subsection{Algorithmic Strategies and Application-Level Contexts}
Beyond architectural integration, the choice of decomposition algorithm deeply influences convergence speed and hardware parallelism. Jacobi methods, while iterative, offer excellent orthogonality preservation and are naturally parallelizable — especially when applied via sweep-based scheduling.

Torun et al. \cite{svd_architecture-4} systematically benchmarked Jacobi algorithms across CPU, GPU, and FPGA platforms. Their findings revealed significant memory irregularity on GPUs, resulting in poor cache utilization and warp divergence. FPGA-based designs, in contrast, performed well due to their predictable access patterns and deep pipelining capability. Athi et al. \cite{svd_architecture-5} further refined the Jacobi method by selecting maximal off-diagonal entries per sweep, achieving faster convergence and improved numerical stability — an approach that directly informs the rotation logic of our proposed design.

Beyond architectural innovations, PCA has also been widely applied in various domain-specific systems as a pre-processing step. For example, PCA has been used for odor classification with neural networks \cite{pca_architecture-10}, gas sensor signal processing \cite{pca_architecture-11}, and data compression in wireless sensor networks \cite{pca_architecture-11}. Additionally, Ali et al. \cite{pca_architecture-11} proposed a hardware PCA implementation for gas identification using high-level synthesis on a Zynq SoC platform. While these works demonstrate PCA’s utility across domains, they are either application-driven or use high-level abstractions, and thus do not contribute to the architectural questions we address in this paper.

\subsection{Summary and Motivation}
While a diverse set of designs target individual stages of PCA acceleration, there remains a lack of a cohesive architecture that combines high-throughput matrix multiplication, pipelined decomposition, and memory-aware data reuse in a fully scalable form. Existing solutions either operate on small matrices, omit key PCA stages, or depend on high-level synthesis and third-party IPs that limit control and scalability.

To address these limitations, we propose \textit{MANOJAVAM} — a unified, fully RTL-programmed architecture that tightly integrates systolic matrix multiplication with a pipelined Jacobi SVD engine. Our design introduces block streaming, operand-mode-aware caching, and RTL-based sweep scheduling for parallel CORDIC-based rotations, all within a reconfigurable fabric. This combination of modularity, performance, and scalability sets \textit{MANOJAVAM} apart from all prior works.
\section{Principal Component Analysis}
Principal Component Analysis (PCA) is a method for dimensionality reduction that maps high-dimensional data onto a low-dimensional subspace, retaining maximum variance. It proceeds via the identification of orthogonal Principal Components (PCs), each formed as a linear combination of the original correlated features\cite{pca_review_1, pca_review_2}. PCA is described in Algorithm.\ref{alg:pca}.   

\begin{algorithm}
\caption{Principal Component Analysis (PCA)}
\label{alg:pca}
\begin{algorithmic}[1]
\REQUIRE Dataset $X \in \mathbb{R}^{M \times N}$
\ENSURE Projected data $O \in \mathbb{R}^{M \times K}$

\STATE Standardize $X$
\STATE Compute covariance matrix $C \gets X^\top X$
\STATE Compute eigenvalues and eigenvectors of $C$
\STATE Select top $K$ components using EVCR or Scree plot
\STATE Project: $O \gets X V_K$
\RETURN $O$
\end{algorithmic}
\end{algorithm}

The initial step in the PCA pipeline involves dataset standardization, as defined in (\ref{Equation : Standardization of input dataset}), to ensure that features having large numerical ranges do not disproportionately bias output principal components\cite{pca_review_1}. In the \textit{MANOJAVAM} architecture, the input dataset $X$ is assumed to be pre-standardized to zero mean and unit variance. This is done so that \textit{MANOJAVAM} can entirely dedicate its datapath to accelerating covariance matrix computation and SVD, which are intensive operations and have time complexities of $O(n \cdot d^{2})$ and $O(d^{3})$ respectively. 

\begin{equation}
    y_{i} = \frac{x_{i} - \mu}{\sigma_{j}}
    \label{Equation : Standardization of input dataset}
\end{equation}

Post standardization, the covariance matrix is constructed using equation (\ref{Equation : Covariance Matrix Computation}). While $C$ is symmetric, \textit{MANOJAVAM} is designed to compute the full $N \times N$ covariance matrix to avoid complex control logic associated with computing only the upper or lower triangular matrix and reflecting the same to generate the full matrix.

\begin{equation}
    C = X^{T}X
    \label{Equation : Covariance Matrix Computation}
\end{equation}

Following covariance matrix computation, the eigendecomposition is performed to extract orthogonal principal components. \textit{MANOJAVAM} implements the Cyclic Jacobi Method for SVD acceleration due to its high degree of parallelism and suitability for fixed point precision\cite{jacobi_algo_3, jacobi_algo_4}. The hardware realization of this stage involves a specialized pipelined Jacobian Unit. First, a Data Lookup Engine (DLE) streams in the output covariance matrix to obtain the maximum off diagonal element $c_{pq}$ and its corresponding diagonal elements $c_{pp}$ and $c_{qq}$. In order to achieve this through streaming in the covariance matrix at runtime, the DLE uses a combination of linear scan with tile aware filtering to obtain the maximum off diagonal element and corresponding diagonal elements on the fly. This approach helps avoid BRAM reads and passes to obtain the same elements, degrading performance. These elements are then streamed into the CORDIC Engine, which computes the rotation angle $\theta$ and subsequent trigonometric transformations. By decomposing the $d \times d$ eigendecomposition problem into a sequence of local $2 \times 2$ Givens rotations, the architecture maintains high numerical stability in fixed-point arithmetic.  

Upon the convergence of the Jacobi sweeps, the diagonal elements of the transformed covariance matrix represent the eigenvalues $\lambda$, which quantify the variance captured by each corresponding Eigenvector. To achieve dimensionality reduction, only a subset of these components are retained. The optimal number of retained components, denoted as \textit{k}, is determined by analyzing the relative contribution of each Eigenvalue to the total variance. The choice of $k$ can be determined easily using a combination of the Explained Variance Contribution Ratio (EVCR) and Cumulative Variance Contribution Ratio (CVCR). EVCR measures the individual contribution of the $i^{th}$ Eigenvector to the pool of Eigenvectors, and CVCR quantifies the total variance retained when mapping the data into the space encompassed by the first $k$ principal components. The equation for EVCR and CVCR is given in (\ref{Equation : EVCR}) and (\ref{Equation : CVCR}).

\begin{equation}
    EVCR = \frac{\lambda_{i}}{\sum_{m=1}^{N} \lambda_{m}}
    \label{Equation : EVCR}
\end{equation}
    
\begin{equation}
    CVCR = \frac{\sum_{i=1}^{k} \lambda_{i}}{\sum_{m=1}^{N} \lambda_{m}}
    \label{Equation : CVCR}
\end{equation}

Where $\lambda_{i}$ denotes the $i^{th}$ Eigenvalue obtained.

Another method frequently employed is the Scree plot, in which the Eigenvalue index is plotted along the X-axis, and the Eigenvalue along the Y-axis. The plot is always a decreasing plot. The rule of thumb is to usually consider one less of all the Eigenvalues to the left of the inflection point in the curve, although this must be combined with EVCR or CVCR for robust analysis. 

Finally, the original dataset is then projected onto the new subspace. This is achieved after the eigenvectors are arranged in columns, and the original dataset matrix is multiplied by the Eigenvector matrix. (\ref{Equation : Projection}) depicts the projection.

\begin{equation}
    O_{m \times k} = X_{centered}V_{k} \in \mathbb{R}^{M \times K}
    \label{Equation : Projection}
\end{equation}
\section{Computational Bottlenecks Associated with PCA}
The primary challenge in scaling Principal Component Analysis (PCA) for high-dimensional, real-time datasets lies in the asymmetric growth of its core computational stages. While the covariance matrix computation scales at $O(n \cdot d^{2})$, the subsequent eigendecomposition (via the Cyclic Jacobi method) requires $O(d^{3})$ operations per sweep. Since the Jacobi method typically converges within a constant number of sweeps, the $O(d^{3})$ complexity dominates only when the feature dimension $d$ is large relative to the sample size $n$.

\begin{figure*}[ht]
    \centering
    \subfloat[]{
        \includegraphics[width=0.47\linewidth]{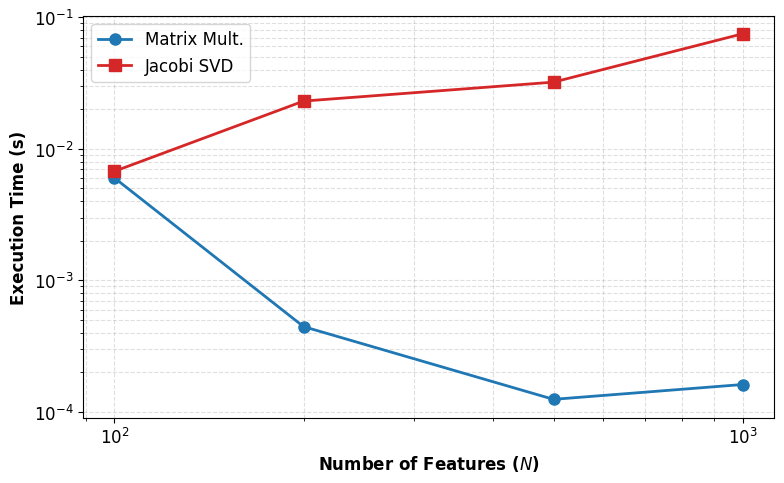}
        \label{fig:bottleneck_rows}
    }
    \hfill
    \subfloat[]{
        \includegraphics[width=0.47\linewidth]{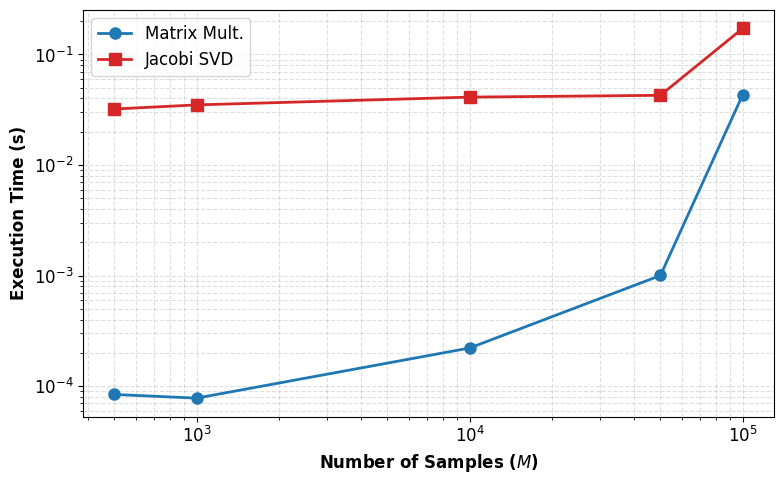}
        \label{fig:bottleneck_features}
    }
    \caption{Breakdown of PCA execution time into matrix multiplication and SVD components under different dataset dimensions. (a) SVD dominates when the number of features increases. (b) Matrix multiplication dominates as the number of rows increases.}
\end{figure*}

To establish a high-performance baseline, the computational bottlenecks of the PCA pipeline were profiled on an NVIDIA RTX A6000 GPU (Ampere architecture, 48GB GDDR6 VRAM). As illustrated in Fig. \ref{fig:bottleneck_rows} and Fig. \ref{fig:bottleneck_features}, the execution time was analyzed across two distinct scaling regimes.

In the constant row regime (Fig. \ref{fig:bottleneck_rows}), it is observed that as the number of features in the dataset $d$ approaches 1000, the $O(d^3)$ complexity of the Jacobi rotations begin to dominate the total latency. Conversely, in the constant feature regime (Fig. \ref{fig:bottleneck_features}), increasing the datapoints $n$ to $10^5$ shifts the bottleneck towards covariance matrix computation. 

These observations necessitate the need for an accelerator to accelerate covariance matrix computation and SVD for PCA. Unlike GPUs that suffer from non-linear scaling and software-stack overheads, our unified FPGA design provides a deterministic, pipelined dataflow that accelerates both matrix multiplication and iterative rotations in a single, high-throughput fabric.
\section{The Jacobian Algorithm}
The Jacobi algorithm is widely employed to compute the eigendecomposition of the covariance matrix. When compared to the Golub-Kahan algorithm\cite{jacobi_algo_1, jacobi_algo_2}, the Jacobi algorithm presents the following advantages: It is highly parallelizable, results in symmetric memory access patterns, and supports fixed point implementations\cite{jacobi_algo_3}. Furthermore, it avoids the hardware-intensive divisions and square roots required by the QR methods, replacing them with area-efficient CORDIC micro operations\cite{cordic_1, cordic_2}. The Jacobi algorithm is presented in Algorithm.\ref{alg:jacobi-svd}.

\begin{algorithm}
\caption{Jacobi Eigenvalue Algorithm}
\label{alg:jacobi-svd}
\begin{algorithmic}[1]
\REQUIRE Symmetric matrix $C \in \mathbb{R}^{N \times N}$
\ENSURE Diagonalized matrix $C$, eigenvectors in $V$

\STATE Initialize $V \gets I$
\WHILE{off-diagonal norm $> \varepsilon$}
    \STATE Find indices $(p, q)$ such that $|c_{pq}|$ is maximum
    \STATE Compute $\theta \gets \frac{1}{2} \tan^{-1}\left( \frac{2c_{pq}}{c_{pp} - c_{qq}} \right)$
    \STATE Set $\cos\theta$ and $\sin\theta$ accordingly
    \STATE Construct rotation matrix $R$ as identity with:
        \STATE \hskip1em $R_{pp} \gets \cos\theta$, $R_{qq} \gets \cos\theta$
        \STATE \hskip1em $R_{pq} \gets \sin\theta$, $R_{qp} \gets -\sin\theta$
    \STATE Update $C \gets R^\top C R$
    \STATE Update $V \gets V R$
\ENDWHILE
\RETURN $C$, $V$
\end{algorithmic}
\end{algorithm}

The algorithm proceeds by obtaining the largest off-diagonal element $c_{pq}$ of the covariance matrix $C$ and the corresponding indices $p$ and $q$. The rotation angle, $\theta$, is then calculated employing (\ref{Equation : Jacobian Rotation Angle}).
\begin{equation}
    \theta = \frac{1}{2} \tan^{-1} \left( \frac{2c_{pq}}{c_{pp} - c_{qq}} \right)
    \label{Equation : Jacobian Rotation Angle}
\end{equation}

The rotation angle $\theta$ reorients the coordinate system such that element $c_{pq}$ becomes zero, resulting in the covariance matrix moving closer to its diagonal form. The Data Lookup Engine (DLE) executes a pivoting strategy by identifying the maximum off-diagonal element ($a_{i,j}$) for each rotation. While the CORDIC units wait for these indices, this approach ensures that each iteration achieves the maximum reduction in off-diagonal energy. 

The rotation matrix $R$ is arrived at by initializing an identity matrix, and populating $R_{pp} = cos\theta$, $R_{pq}=sin\theta$, $R_{qp} = -sin\theta$ and $R_{qq} = cos\theta$ as shown in (\ref{Equation : rotational matrix initialisation}).
\begin{equation}
R =
\begin{pmatrix}
1      & 0      & \cdots &        &        & \cdots & 0 \\
0      & 1      & \cdots &        &        & \cdots & 0 \\
\vdots & \vdots & \ddots &        &        &        & \vdots \\
       &        &        & \cos \theta & \sin \theta &        &        \\
       &        &        & -\sin \theta & \cos \theta &        &        \\
\vdots & \vdots &        &        &        & \ddots & \vdots \\
0      & 0      & \cdots &        &        & \cdots & 1
\end{pmatrix}
\label{Equation : rotational matrix initialisation}
\end{equation}

The Jacobi algorithm is grounded in a similarity transformation framework, which expresses the diagonalization of a matrix A through a sequence of orthogonal transformations, as shown in (\ref{Equation : similarity transformation theorem}).
\begin{equation}
    A^{'} = P^{-1}AP
    \label{Equation : similarity transformation theorem}
\end{equation}

When the transformation matrix P is orthogonal (ie., $P^{-1} = P^{T}$), this reduces to the Spectral theorem, as described in (\ref{Equation : spectral theorem}).
\begin{equation}
    A^{'} = P^{T}AP
    \label{Equation : spectral theorem}
\end{equation}

Since the covariance matrix $C$ is real and symmetric, the spectral theorem applies directly, enabling its diagonalization via orthogonal similarity transformations, depicted in (\ref{Equation : Jacobi Rotation Equation}).
\begin{equation}
    C^{'} = R^{T} C R
    \label{Equation : Jacobi Rotation Equation}
\end{equation}

The Jacobi algorithm applies a sequence of such rotations—referred to as Jacobi sweeps—to iteratively zero out off-diagonal elements. The method exhibits quadratic convergence, whereby each sweep significantly reduces the magnitude of the largest off-diagonal entry. 

The convergence of the Cyclic Jacobi method is typically monitored using the off-diagonal Frobenius norm, defined as:
\begin{equation}
    E_{off}(A) = \sqrt{\sum_{i=1}^{n} \sum_{j \neq i}^{n} |a_{ij}|^2}
\end{equation}
where $a_{ij}$ represents the off-diagonal elements of the evolving covariance matrix. In many software-based SVD solvers, an 'early-exit' strategy is employed where the algorithm terminates once $E_{off}$ falls below a predefined threshold $\epsilon$.

However, implementing a real-time Frobenius norm unit on-chip presents significant hardware challenges. The calculation requires a high-precision Square-Root-of-Sum-of-Squares (SRSS) pipeline across the entire $N \times N$ matrix, which introduces substantial logic overhead, routing congestion, and a potential reduction in the maximum clock frequency ($F_{max}$). 

To maintain the high-throughput, streaming nature of \textit{MANOJAVAM}, we move the convergence analysis offline. By performing a comprehensive Frobenius norm study across varied data modalities, we determine a fixed iteration count ($i$) that guarantees convergence. The fixed iteration count is also made large enough to accommodate ill-conditioned data at the input of the accelerator. This approach eliminates the need for complex monitoring logic while ensuring a deterministic execution latency.
\section{Architecture of Manojavam}
\begin{figure*}[t]
    \centering
    \includegraphics[width=1\linewidth]{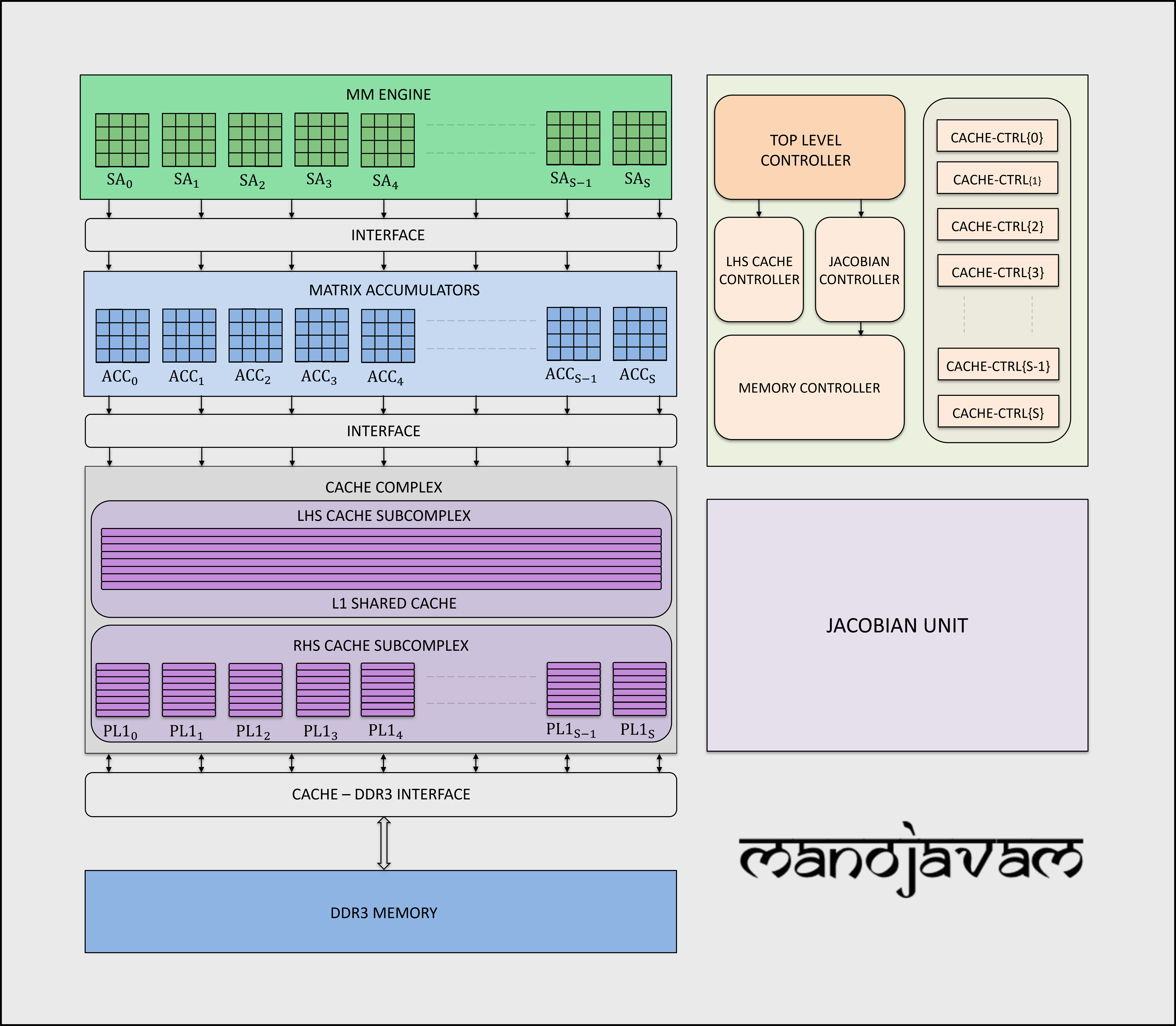}
    \caption{Manojavam : High Level Architecture}
    \label{fig:manojavam high level architecture}
\end{figure*}

This section introduces \textit{MANOJAVAM}, a hardware accelerator, to speed up matrix multiplication and SVD for PCA. The core of \textit{MANOJAVAM}'s architecture is the Matrix Multiplication Engine (MM-Engine), which integrates an array of $T$x$T$ systolic arrays alongside dedicated matrix accumulators. This unit performs both covariance matrix computation and rotations by employing block streaming to scale to large input matrices. Eigendecomposition is enabled using a lightweight Jacobian Unit, which identifies the maximum off-diagonal entry of the covariance matrix, and employs CORDIC-based micro-operations to calculate the parameters of the Given's rotation matrix. Rotations are carried out in the MM-Engine, eliminating the need for a separate rotation datapath, and reducing hardware redundancy. A two-level cache hierarchy ensures optimized operand delivery, and a coordinated hierarchy of RTL controllers ensures synchronization across all datapath units. The high-level architecture of \textit{MANOJAVAM} is shown in Fig.\ref{fig:manojavam high level architecture}. Central to its efficiency is a unified data-path that treats the covariance computation and eigendecomposition not as separate tasks, but as two operational modes of the same high-throughput fabric.

\subsection{Matrix Multiplication Accelerator Architecture}
\textit{MANOJAVAM}’s matrix multiplication accelerator adopts a block streaming methodology. Partial product submatrices are computed in parallel and accumulated across corresponding row and column blocks to form the complete matrix product\cite{pca_architecture-3}.This streaming strategy guides the architectural design of the matrix multiplication core, comprising an array of compact $T \times T$ systolic arrays. To mitigate the memory wall encountered by general-purpose GPUs (as demonstrated in Section IV), \textit{MANOJAVAM} utilizes a deterministic block-streaming methodology. By partitioning large $N \times D$ matrices into $T \times T$ tiles, the architecture maintains a constant memory footprint regardless of the sample count, ensuring that the system avoids the latency penalties associated with cache thrashing. Each systolic array in the MM-Engine is responsible for computing a submatrix of the final product matrix. Compared to a monolithic Tensor Processing Unit (TPU) core, this modular approach provides several key advantages: support for arbitrary matrix dimensions, efficient handling of boundary cases, improved resource utilization, and reduced power consumption via selective unit-level gating. In contrast, a large TPU becomes suboptimal when input matrix dimensions are misaligned with its array configuration, leading to under-utilization with small matrices and complex batching strategies for larger inputs\cite{systolic_arrays_1}. The modular approach allows for efficient processing of rectangular matrices, commonly seen in PCA workloads. 

\begin{figure}
    \centering
    \includegraphics[width=1\linewidth]{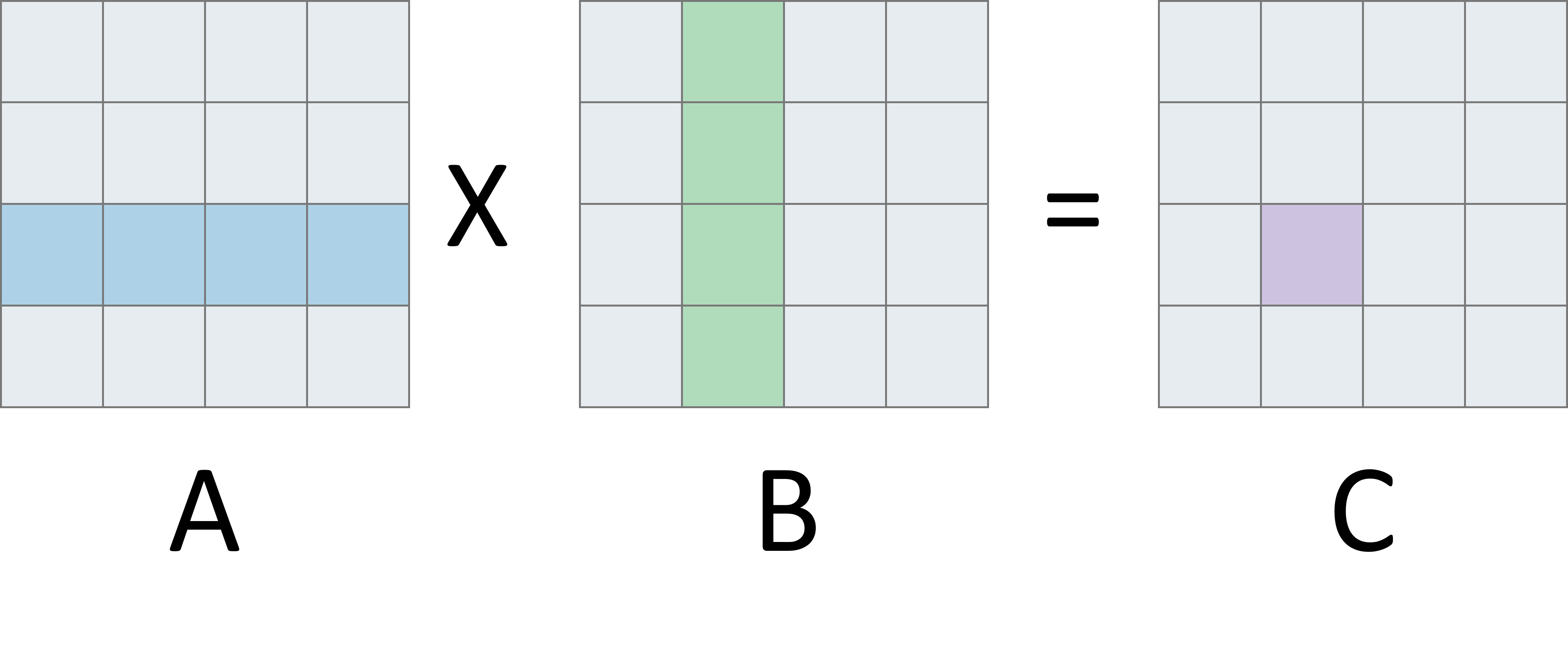}
    \caption{Block Streaming Illustration}
    \label{fig:block streaming illustration}
\end{figure}

Block streaming is highly effective when dealing with very large matrix operands that cannot be streamed in all at once into the core computing units. A "block" of tiles is streamed sequentially to compute the partial product tiles. These tiles are then accumulated to create the output tile\cite{block_stream_1}. An illustration of block streaming is shown in Fig.\ref{fig:block streaming illustration}.

\begin{figure}
    \centering
    \includegraphics[width=1\linewidth]{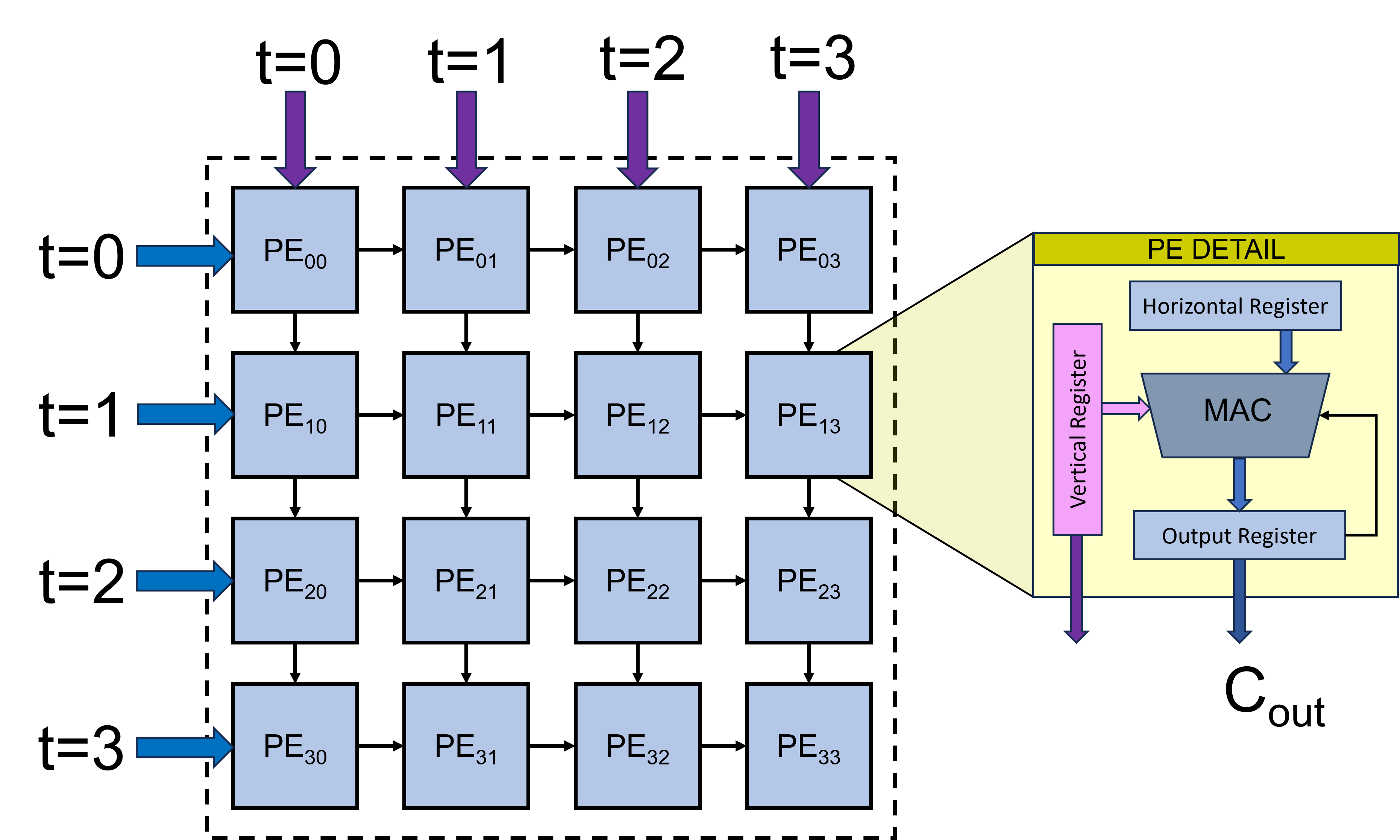}
    \caption{Systolic Array}
    \label{fig:systolic array}
\end{figure}

Systolic arrays are widely adopted for implementing matrix multiplication in hardware due to their highly regular, pipelined structure, enabling efficient spatial dataflow and minimal control overhead. Their inherent regularity allows for compact layouts and predictable timing, making them well-suited for FPGA and ASIC implementations. Systolic architectures have been employed in a variety of high-performance systems, most notably in Google’s Tensor Processing Unit (TPU)\cite{systolic_arrays_2} and several other prior work on Machine Learning accelerators, DNN Accelerators and Digital Signal Processing\cite{systolic_arrays_3, systolic_arrays_4, systolic_arrays_5, systolic_arrays_6, systolic_arrays_7, systolic_arrays_8, systolic_arrays_9, systolic_arrays_10, systolic_arrays_11, systolic_arrays_12, systolic_arrays_13}. Compared to traditional row-column multipliers, systolic arrays offer high parallelism, better resource utilization, and improved scalability for large matrix operations. As depicted in Fig.\ref{fig:systolic array}, input operands are streamed into the systolic array in a systolic fashion, as suggested by the parallelogram input data profile to the systolic array. Each element of the systolic array is a Multiply and Accumulate (MAC) unit. Data is fed into these Processing Elements (PEs) in a skewed systolic fashion, allowing for $100\%$ functional occupancy of the MAC units once the pipeline is primed. This spatial dataflow is architecturally significant as it minimizes global wire congestion and reduces power consumption by localizing data movement strictly between adjacent PEs, avoiding the high-energy cost of global bus transactions.

Matrix multiplication is performed over multiple passes, with the number of passes determined by the dimensions of the input operand matrices. The MM-Engine in \textit{MANOJAVAM} consists of $S$ $T$x$T$ systolic arrays, enabling up to $S$ matrix multiplication operations to be parallelized concurrently. Each systolic array is responsible for computing a specific submatrix of the overall matrix product, corresponding to a pair of row and column blocks. These row and column blocks are further partitioned into $T$x$T$ tiles that align with the dimensions of the systolic arrays. This tile-based partitioning supports high-throughput block streaming and ensures that data is processed in a spatially and temporally efficient manner. For instance, in computing the covariance matrix $C$, the submatrix $C_{00}$ is calculated by the systolic array $SA_{0}$, which accumulates the products of row block-0 of $X^{T}$ and column block-0 of $X$. Each systolic array is paired with a dedicated accumulator to maintain correct accumulation of partial products across passes. This preserves numerical consistency and prevents accumulation of the incorrect matrix product from the MM-Engine.

First, the tiles are queried from the shared LHS and private RHS caches associated with the systolic array-matrix accumulator pair. The tiles first pass through Matrix Padding Units (MPU) at the interface between the MM-Engine and the caches. The MPU aids in inputting the tiles in a systolic fashion. There are two separate MPUs developed for the two input operands, as they have different systolic profiles. After the formation of matrix products, the output from the systolic array is fed into the matrix accumulator. This accumulates the $T$x$T$ product tiles associated with the row and column block. Upon completion of all row and column tile iterations, the top-level controller signals the end of partial product submatrix computation. The output from the matrix accumulator is then forwarded both to the Jacobi Unit for eigendecomposition and to memory for covariance matrix storage.

The MM-Engine is reused for performing matrix multiplication during rotations. Crucially, the MM-Engine is not idle during the eigendecomposition phase; it is repurposed to apply the calculated Givens rotations to the entire covariance matrix. By acting as a parallel transformation engine that updates multiple rows and columns simultaneously, \textit{MANOJAVAM} eliminates hardware redundancy and maximizes the area-efficiency of the silicon footprint. To manage this distinction, the top-level controller issues a one-bit mode signal to the datapath, indicating whether the system is performing covariance computation or rotation at a given time. This ensures that the rest of the datapath executes the correct operation while appropriately scheduling operands for matrix multiplication during both covariance computation and rotation stages. This allows the reuse of computational resources of the accelerator across both stages.  

\subsubsection{Illustration}
To illustrate the working of the matrix multiplication accelerator, assume an accelerator configuration of $T=4$ and $S=8$. Consider the input dataset to be of size 1000x1024, ie-1000 rows and 1024 features. If this is represented by matrix $X$, then $X$ is of dimension 1000x1024, and $X^{T}$ is of dimension 1024x1000. Thus, the number of row blocks in $X^{T}$ is $1024/4 = 256$, and the number of column blocks in $X$ is again $1024/4 = 256$. Each row and column block is divided into $1000/4 = 250$ tiles, each of dimension 4x4. In pass 1, the 8 systolic arrays are scheduled as follows - $SA_{0}$ is associated with the computation of partial product matrix $R_{0}C_{0}$, $SA_{1}$ is associated with the computation of partial product matrix $R_{0}C_{1}$, $SA_{2}$ is associated with the computation of partial product matrix $R_{0}C_{2}$ and similarly, $SA_{7}$ is associated with the computation of partial product matrix $R_{0}C_{7}$. This is held on until all 256 tiles associated with $R_{0}$, and each of the 256 tiles associated with $C_{0}$ to $C_{7}$, are scheduled and their output product tiles are computed and forwarded to the matrix accumulators indexed with the systolic array. After all tiles in the row block and all of the column blocks have been scheduled, the top-level controller asserts that the 8 partial product submatrices have been successfully computed, and it arranges for the subsequent row blocks and column blocks to be fed in. As there are 256 column blocks to be computed, the top level controller schedules $C_{8}$ to $C_{15}$ for computation, while retaining $R_{0}$. The process continues until all of the 256 column blocks have been passed over for $R_{0}$. This ensures that one row of the output covariance matrix is completed, and continues until all the row blocks are iterated.

\subsection{Cache Subsystem}
\textit{MANOJAVAM} employs a two-tier cache hierarchy system tailored for efficient matrix operand delivery. Operand A (LHS) is served by a single shared cache while operand B (RHS) is distributed across $S$ private caches. Each of the private caches is tightly coupled to its corresponding systolic array and matrix accumulator. In the architecture, the shared and private caches are directly mapped. This cache system organization is architecturally driven - operand A is broadcasted and reused across multiple passes across systolic arrays, while operand B is unique to each array instance. These data access patterns have led to the shared and localized caching to ensure efficient operand delivery. This asymmetric cache architecture is specifically engineered to facilitate high efficiency block streaming. By utilizing a shared LHS cache, the architecture performs a single 'broadcast' read, serving all processing elements simultaneously and reducing global memory bandwidth requirements by a factor of $S$. Conversely, each systolic array is equipped with a Private RHS Cache to store the distinct vertical tiles required for its specific submatrix computation. 

Unlike conventional cache designs that store matrix rows directly, \textit{MANOJAVAM} cache store complete matrix tiles, each encompassing a $T \times T$ submatrix in a single row entry. This layout avoids the inefficiency of issuing $T$ separate cache reads to reconstruct one tile. Instead, a tile is fetched entirely in a single burst read to cache memory. An offline script flattens the raw input dataset into the proposed layout and then loaded into memory at system reset. The cache rows are populated upon cache misses.

Each cache block is managed by a cache controller, totaling the number of controllers across the architecture to $(S+1)$. These cache controllers independently issue address instructions for operand data from memory. These controllers dynamically adjust their cache write-miss policies based on the phase of PCA computation. When computing the covariance matrix, $C=X^{T}X$, the access pattern resembles the Livermore loop benchmark\cite{cache_subsystem_1}, given in (\ref{eq:Livermore benchmark equation}). Hence, during the computation of covariance matrix, the caches must operate with the write-around policy for cache misses to ensure no cache pollution and high performance\cite{cache_subsystem_3}.

\begin{equation}
    A[j] = B[j] + C[j]
    \label{eq:Livermore benchmark equation}
\end{equation}

In contrast, during the rotation phase, ie- $C^{'} = R^{T}CR$ and $V = VR$,  the access pattern aligns with the SAXPY-like behavior in the Linpack benchmark\cite{cache_subsystem_2}, given in (\ref{eq:Linpack benchmark equation}), where write-allocate no-fetch-on-write policy is more appropriate\cite{cache_subsystem_3}.
\begin{equation}
    A[j] = A[j] + B[j]
    \label{eq:Linpack benchmark equation}
\end{equation}

A mode signal propagates through the system to reconfigure cache controller behavior at runtime, enabling seamless adaptation between these two execution modes and ensuring policy coherence with workload characteristics.

\subsection{Jacobian Unit}
\begin{figure*}[h]
    \centering
    \includegraphics[width=1\linewidth]{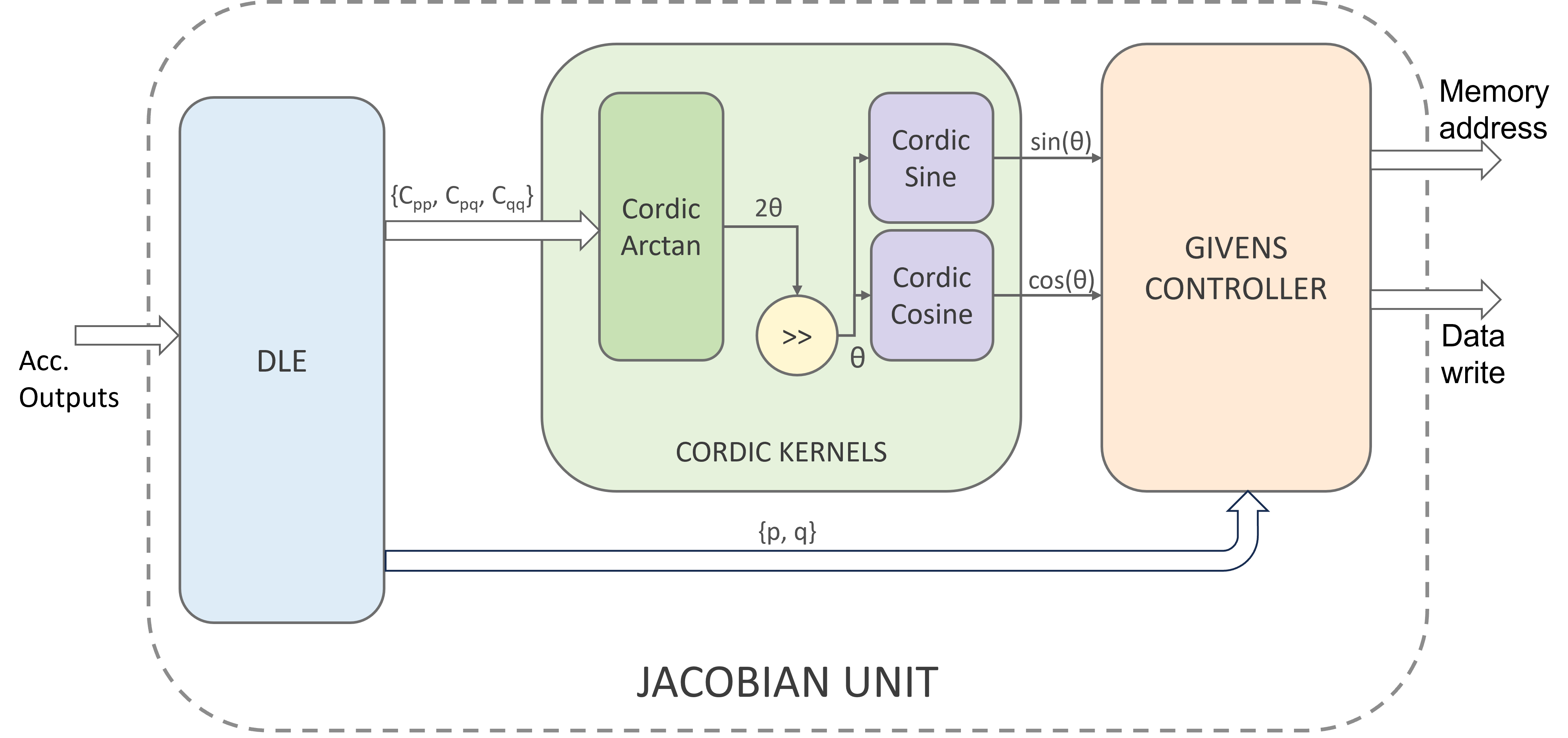}
    \caption{Jacobian Unit Architecture}
    \label{fig:jacobian unit architecture}
\end{figure*}

The Jacobian Unit in the accelerator serves as the bridge between the covariance matrix and the computation of the Givens rotation matrix, $R$. The Givens matrix is employed in rotations to converge at the eigenvalues and eigenvectors of the input covariance matrix. The high level design of the Jacobian unit is given in Fig. \ref{fig:jacobian unit architecture}.

The Jacobian unit comprises of the Data Lookup Engine (DLE), which interfaces directly with the output ports of the systolic arrays. The DLE streams in the covariance matrix directly from the accumulation stage and uses a linear scan to identify the maximum off-diagonal element $c_{pq}$, corresponding indices $(p, q)$ and elements $c_{pp}$ and $c_{qq}$ for the CORDIC engine. By resolving the pivot coordinates at the hardware interface, the DLE removes the need for redundant memory round-trips to the BRAM, providing a low-latency transition to the rotation phase.

To maintain mathematical validity during the search, the Jacobian Controller implements tile-aware filtering.  By monitoring the current row-block index, the DLE selectively masks the main diagonal elements ($c_{pp}$) during the comparison phase, ensuring the engine only locks onto valid off-diagonal candidates. 

On reset, a global register is initialized to track the maximum off-diagonal element and its associated diagonal terms $c_{pp}$ and $c_{qq}$. As each matrix multiplication pass produces $S$ partial results from the accumulators, the Jacobian Controller inspects the current row block index and selectively filters out diagonal elements accordingly. For instance, during the processing of row block $R_{0}$, the diagonal elements from $Acc_{0}$ (which correspond to the main diagonal of $C$) are discarded, while all elements from accumulators $Acc_{1}$–$Acc_{S}$ are considered. In subsequent passes within $R_{0}$, all elements are valid for checking. A similar rule applies for row block $R_{1}$, where only $Acc_{1}$ diagonals are excluded. This localized, tile-aware filtering ensures that only valid off-diagonal candidates are passed to the DLE. Once all checks are completed for a given row block, the controller outputs the values of $c_{pq}$, $c_{pp}$ and $c_{qq}$, along with their corresponding indices $p$ and $q$, for use in the Givens rotation step.

The outputs from the DLE—specifically the values $c_{pq}$, $c_{pp}$ and $c_{qq}$, are forwarded to the CORDIC kernel\cite{cordic_1, cordic_2}, which computes the sine and cosine of the rotation angle $\theta$ required for the Jacobi transformation. The rotation angle is computed using the relation in (\ref{Equation : Jacobian Rotation Angle}). This is implemented via a pipelined CORDIC arctangent unit, followed by a 1-bit right shifter. The resulting angle is then fed to additional CORDIC units that compute $sin(\theta)$ and $cos(\theta)$ in parallel. These trigonometric values are then passed to the Givens Controller, which initializes the corresponding entries of the Givens rotation matrix in memory. This matrix is pre-initialized to the identity matrix upon reset, and subsequently updated at runtime. Since the Givens matrix is reused in multiple operations, a write-back to memory is necessary and architecturally justified. The Givens Controller uses the indices $p$ and $q$ to identify the appropriate memory locations, and updates them in a single event using a dual-port memory system.

Once written, the top-level controller signals the beginning of the rotation phase by asserting the $mode$ signal. Although initial accesses to the rotation matrix result in cache misses, the tiles are populated in the LHS and RHS operand caches.
\section{Experimental Setup and Results}
\subsection{FPGA Implementation}
\textit{MANOJAVAM} is developed entirely in synthesizable Verilog (IEEE Std 1364-2005) using the AMD Vivado 2024.1 design suite. The design is implemented on two platforms - a Xilinx Artix-7 xc35atcpg236-1 FPGA and Xilinx Virtex Ultrascale+ xcu250-figd2104-2L-e FPGA. The platforms are chosen to showcase design's deployability across both resource-constrained and resource expansive targets. The ample resources in the Virtex Ultrascale+ FPGA help support the conduct of ablation studies for different tile sizes ($T$) and parallelism indices ($S$). Functional correctness was validated using custom simulation testbenches, while Xilinx Design Constraints (XDC) were applied to guide placement and routing through floorplanning.

Upon implementation, the following utilization report estimates were obtained, and are depicted in Table.\ref{tab:manojavam fpga resource utilization artix-7} and Table.\ref{tab:manojavam fpga resource utilization virtex-ultrascale}.

\begin{table}[ht]
    \centering
    \caption{Manojavam(4,8) FPGA Resource Utilization - Xilinx Artix-7 xc35atcpg236-1}
    \scalebox{1}{
    \begin{tabular}{|c|c|c|c|}
    \hline
    \textbf{Resource} & \textbf{Usage} & \textbf{Total Available} & \textbf{Utilization} \\
    \hline
    LUT & 9796 & 20800 & 47.10 \\
    \hline
    FF & 23077 & 41600 & 55.47 \\
    \hline
    BRAM & 30.5 & 50 & 61 \\
    \hline
    DSP & 64 & 90 & 71.11 \\
    \hline
    \end{tabular}}

    \label{tab:manojavam fpga resource utilization artix-7}
\end{table}

\begin{table}[ht]
    \centering
    \caption{Manojavam(16,32) FPGA Resource Utilization - Xilinx Virtex Ultrascale+ xcu250-figd2104-2L-e}
    \scalebox{1}{
    \begin{tabular}{|c|c|c|c|}
    \hline
    \textbf{Resource} & \textbf{Usage} & \textbf{Total Available} & \textbf{Utilization} \\
    \hline
    LUT & 195814 & 1728000 & 11.33 \\
    \hline
    FF & 143777 & 3456000 & 4.16 \\
    \hline
    BRAM & 940.5 & 2688 & 34.99 \\
    \hline
    DSP & 4096 & 12288 & 33.33 \\
    \hline
    \end{tabular}}

    \label{tab:manojavam fpga resource utilization virtex-ultrascale}
\end{table}

\textit{MANOJAVAM} operates at 200 MHz with 1.271 W power dissipation on the Artix-7 FPGA, and at 434 MHz consuming 16.957 W on the Virtex Ultrascale+ platform.

To provide a conservative performance bound, we developed a cycle-approximate analytical simulator that models a worst-case sequential dataflow. This model accounts for effective access times (EAT) by incorporating a cache hit rate of $p=0.9$ and a $10\times$ penalty for off-chip DRAM access. For a tile size of $T=16$, the simulator calculates the total execution time as the aggregate of data-loading overhead and systolic computation cycles. This ensures that the reported performance metrics represent a strictly attainable lower bound, even under significant memory contention.

The Table.\ref{tab:Comparison against Prior PCA Accelerators} outlines the comparison between prior accelerators and \textit{MANOJAVAM}.

\begin{table*}[htbp]
    \centering
    \caption{Comparison with Prior PCA Accelerators}
    \scalebox{1}{
    \begin{tabular}{|c|c|c|c|c|c|c|c|c|}
    \hline
    \textbf{Architecture} & \textbf{LUT} & \textbf{FF} & \textbf{BRAM} & \textbf{DSP} & \textbf{$F_{max}$ (MHz)} & \textbf{Power (W)} & \textbf{Maximum Dimension (AxB)} \\
    \hline
    Korat.et.al \cite{pca_architecture-1} & 272026 & - & 196 & 2464 & 183 & - & 16x30 \\
    \hline
    Mansoori.et.al \cite{pca_architecture-3} & 47880 & 34048 & 38 & 117 & 90 & 2.37 & 640x480 \\
    \hline
    Shahrouzi.et.al \cite{pca_architecture-5} & 24692 & 49384 & - & 11 & 100 & 1.58 & 64x3823 \\
    \hline
    Wang.et.al \cite{svd_architecture-6} & 13372 & 12723 & - & - & 129.2 & - & 8x8 \\
    \hline
    Das.et.al \cite{pca_architecture-4} & 17971 & 8932  & 74 & - & 139.5 & - & - \\
    \hline
    Torun.et.al \cite{svd_architecture-4} & - & - & -  & - & 178 & 0.26 & 8x8 \\
    \hline
    Fernandez.et.al \cite{pca_architecture-2} & 16045 & 14417 & 28  & - & 124.6 & - & 302500x224 \\
    \hline
    Ma.et.al \cite{svd_architecture-1} & 182605 & - & 3044 & 740 & 183 & - & 128x128 \\
    \hline
    Shiuping.et.al \cite{svd_architecture-2} & 63656 & 63656 & - & - & 130 & - & 8x8 \\
    \hline
    Athi.et.al \cite{svd_architecture-5} & 32368 & 18760 & 48 & 192 & 236 & - & 8x8 \\
    \hline
    Bravo.et.al \cite{pca_architecture-12} & - & - & 40 & 43 & 112.4 & - & 256x256 \\
    \hline
    Kasap.et.al \cite{svd_architecture-7} & 23567 & 21268 & 53 & 64 & 103.88 & - & 4x4  \\
    \hline
    \textbf{\textit{MANOJAVAM} (4,8)} & \textbf{9796} & \textbf{23077} & \textbf{30.5} & \textbf{64} & \textbf{200} & \textbf{1.271} & \textbf{Scale-Invariant}  \\
    \hline
    \textbf{\textit{MANOJAVAM} (16,32)} & \textbf{195814} & \textbf{143777} & \textbf{940.5} & \textbf{4096} & \textbf{434} & \textbf{16.957} & \textbf{Scale-Invariant}  \\
    \hline
    \end{tabular}}

    \vspace{1mm}
    
    \noindent\footnotesize{\textit{Note:} ``–'' indicates that the corresponding figure was not reported in the referenced work.} \\
    \noindent\footnotesize{\textit{Note:} Unlike prior works constrained by fixed on-chip buffer sizes, MANOJAVAM’s block-streaming allows processing of arbitrarily large matrices limited only by external storage capacity.}
    
    \label{tab:Comparison against Prior PCA Accelerators}
\end{table*}

\subsection{Execution Time Analysis on Real World PCA Benchmarks}
This section evaluates the execution time performance of \textit{MANOJAVAM} on a suite of real-world benchmark datasets commonly employed in Principal Component Analysis. The selected datasets summarized in Table.\ref{tab:benchmark_summary} span diverse data modalities and dimension characteristics. These datasets cover a range of application domains including computer vision, hyperspectral imaging, biomedical analysis and text mining.

\renewcommand{\arraystretch}{1.2} 
\begin{table*}[htbp]
    \centering
    \caption{Summary of Benchmark PCA Datasets}
    \scalebox{1}{
    \begin{tabular}{|>{\centering\arraybackslash}p{2.5cm}| 
                    >{\centering\arraybackslash}p{2.5cm}| 
                    >{\centering\arraybackslash}p{2.5cm}| 
                    >{\arraybackslash}p{6.5cm}|} 
    \hline
    \textbf{Dataset} & \textbf{Number of Records} & \textbf{Number of Features} & \textbf{Brief Description} \\
    \hline
    MNIST-8x8\cite{mnist_8x8} & 1797 & 64 & Grayscale images of handwritten digits (8×8 resolution), widely used for digit classification tasks. \\
    \hline
    MNIST-28x28\cite{mnist_28x28} & 70000 & 784 & Standard MNIST dataset with 28×28 grayscale images of digits; high-dimensional image input. \\
    \hline
    CIFAR-10\cite{cifar10} & 60000 & 3072 & RGB images (32×32×3) across 10 object categories; common for low-resolution object recognition. \\
    \hline
    Olivetti Faces\cite{olivetti_faces} & 400 & 4096 & Grayscale face images (64×64 pixels) of 40 individuals with varying facial expressions and lighting. \\
    \hline
    Breast Cancer\cite{breast_cancer_wisconsin} & 45312 & 7 & Biomedical features extracted from mammographic scans for breast cancer diagnosis. \\
    \hline
    20-Newsgroups\cite{20_newsgroups} & 18846 & 1024 & Text documents represented as TF-IDF feature vectors across 20 categories; high-dimensional sparse input. \\
    \hline
    \end{tabular}}
    \label{tab:benchmark_summary}
\end{table*}

\begin{figure*}[h]
    \centering
    \includegraphics[width=0.9\linewidth]{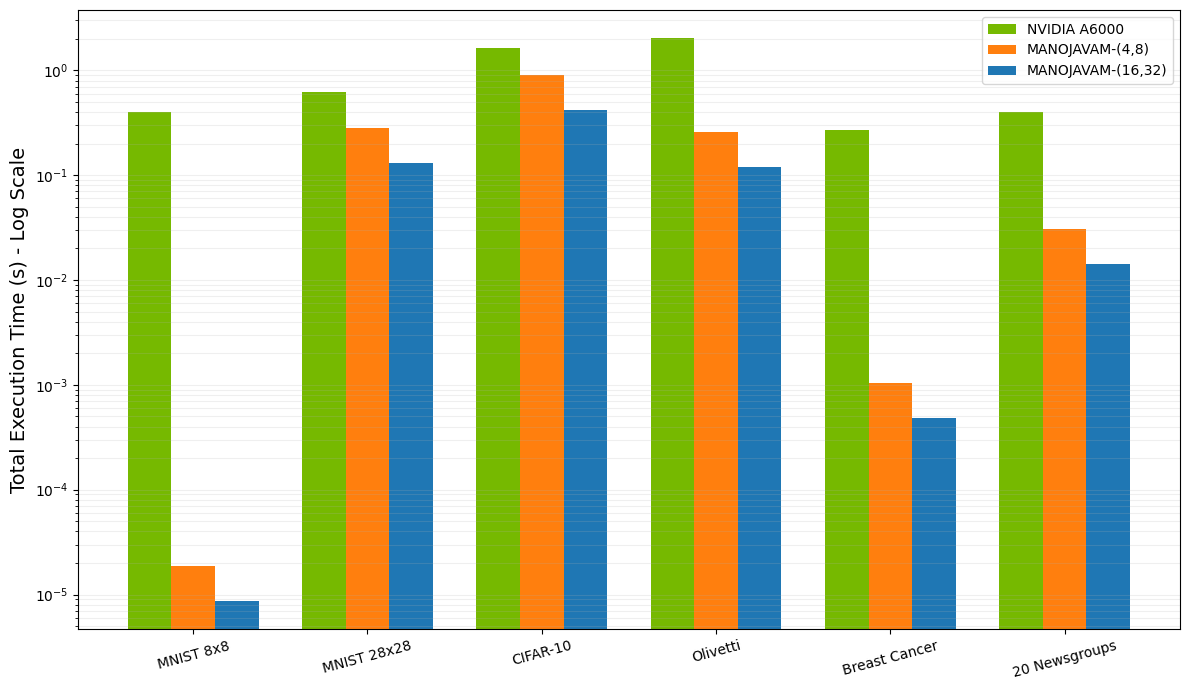}
    \caption{Total Execution Time across Benchmark Datasets profiled across all Platforms}
    \label{fig:pca_benchmarking}
\end{figure*}

For comparisons, \textit{MANOJAVAM} is benchmarked against an NVIDIA A6000 GPU\cite{nvidia2020ampere}. The total execution times for various datasets across all platforms is presented in Fig.\ref{fig:pca_benchmarking}. The results demonstrate that \textit{MANOJAVAM} outperforms the GPU on all datasets. Notably, on the CIFAR-10 dataset which is the most compute intensive benchmark in our dataset suite, \textit{MANOJAVAM} achieves a $3.87\times$ reduction in total execution time compared to the NVIDIA A6000. 

The benchmarking results from the MNIST-8x8 and Breast Cancer datasets show that the GPU performs sub-optimally compared to \textit{MANOJAVAM}. This is attributed to the small sizes of the covariance matrix generated from these datasets which fail to efficiently utilize the massive parallel fabric of the NVIDIA A6000\cite{svd_architecture-4}. The performance bottleneck on the GPU is driven by kernel launch latencies and SIMT (Single Instruction Multiple Thread) branch divergence during iterative Jacobi sweeps. In contrast, \textit{MANOJAVAM} avoids these overheads through dedicated, hard-wired control logic and a parametric tiled dataflow, ensuring high utilization regardless of matrix scale. 

\subsection{Energy Efficiency and Computational Density}
\begin{figure*}[h]
    \centering
    \includegraphics[width=0.9\linewidth]{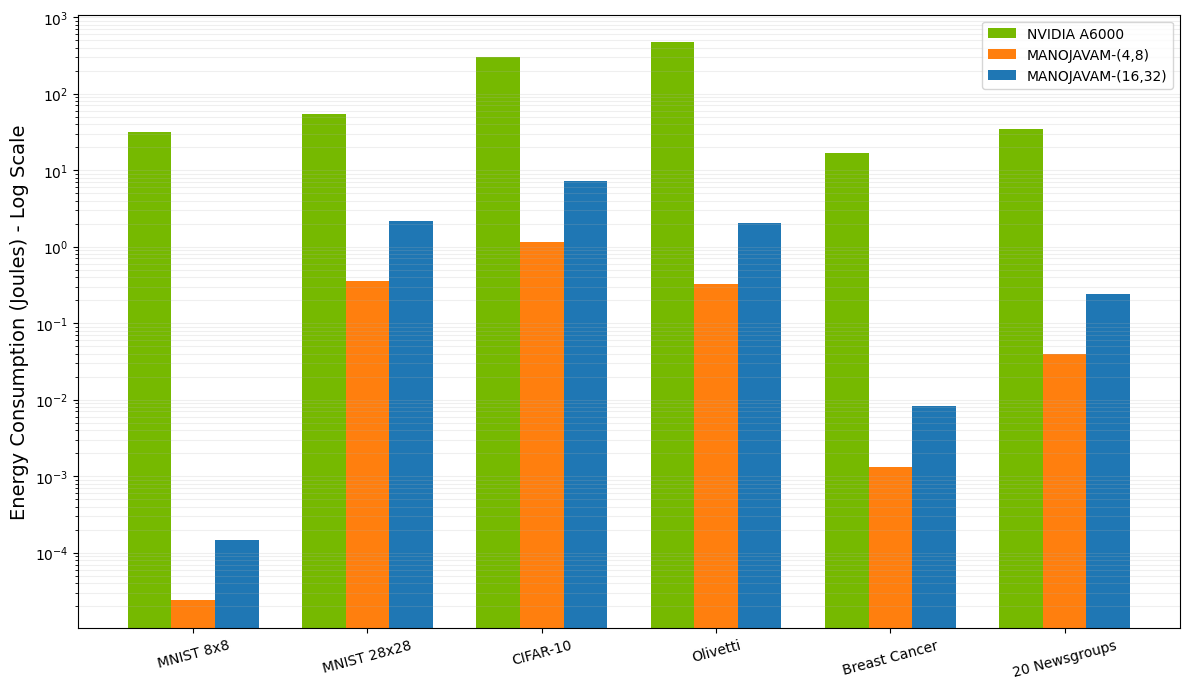}
    \caption{Energy consumption across benchmarks (Log Scale). MANOJAVAM achieves up to $10^6\times$ higher efficiency than the NVIDIA A6000 by eliminating GPU driver overhead and utilizing a deterministic, low-power systolic pipeline.}
    \label{fig:pca_energy_consumption}
\end{figure*}

Fig.\ref{fig:pca_energy_consumption} presents a comparison of the energy profiles for the various datasets across the hardware platforms. The energy consumption ($E$) is defined as the product of peak measured power ($P_{peak})$ and the total end to end execution latency ($T_{total}$). 

The results show that \textit{MANOJAVAM} showcases a significant reduction in energy consumption across all the datasets compared to the NVIDIA A6000. Notably, for smaller datasets such as the MNIST-8x8, the energy gain exceeds five orders of magnitude ($>10^5\times$). This is attributed to a higher power floor in the GPU and CUDA overhead times resulting in higher energy consumption, while the proposed accelerator has a much lower power floor and is purely datapath, resulting in low power operation.

In high-dimensional scenarios such as CIFAR-10 ($N=3072$) and 20 Newsgroups ($N=1024$), where computational kernels typically saturate GPU resources, MANOJAVAM continues to exhibit superior efficiency. The MANOJAVAM-(16,32) configuration, deployed on the Virtex UltraScale+, achieves a $42.14\times$ energy reduction for CIFAR-10 compared to the A6000. This confirms that the proposed systolic Jacobi array and pipelined matrix multiplication units provide higher computational density per Watt than the SIMT-based kernels of modern workstation GPUs.

The results further highlight the advantage of cycle-deterministic dataflow. Unlike the A6000, which exhibits power fluctuations due to dynamic frequency scaling and non-deterministic OS-level interrupts, MANOJAVAM provides a stable, predictable power profile. This makes the proposed accelerator particularly suitable for deployment in mission-critical edge environments where energy budgets are stringent and thermal throttling must be avoided.

\subsection{Frobenius Norm Convergence Analysis}
\begin{figure}[H]
    \centering
    \includegraphics[width=1\linewidth]{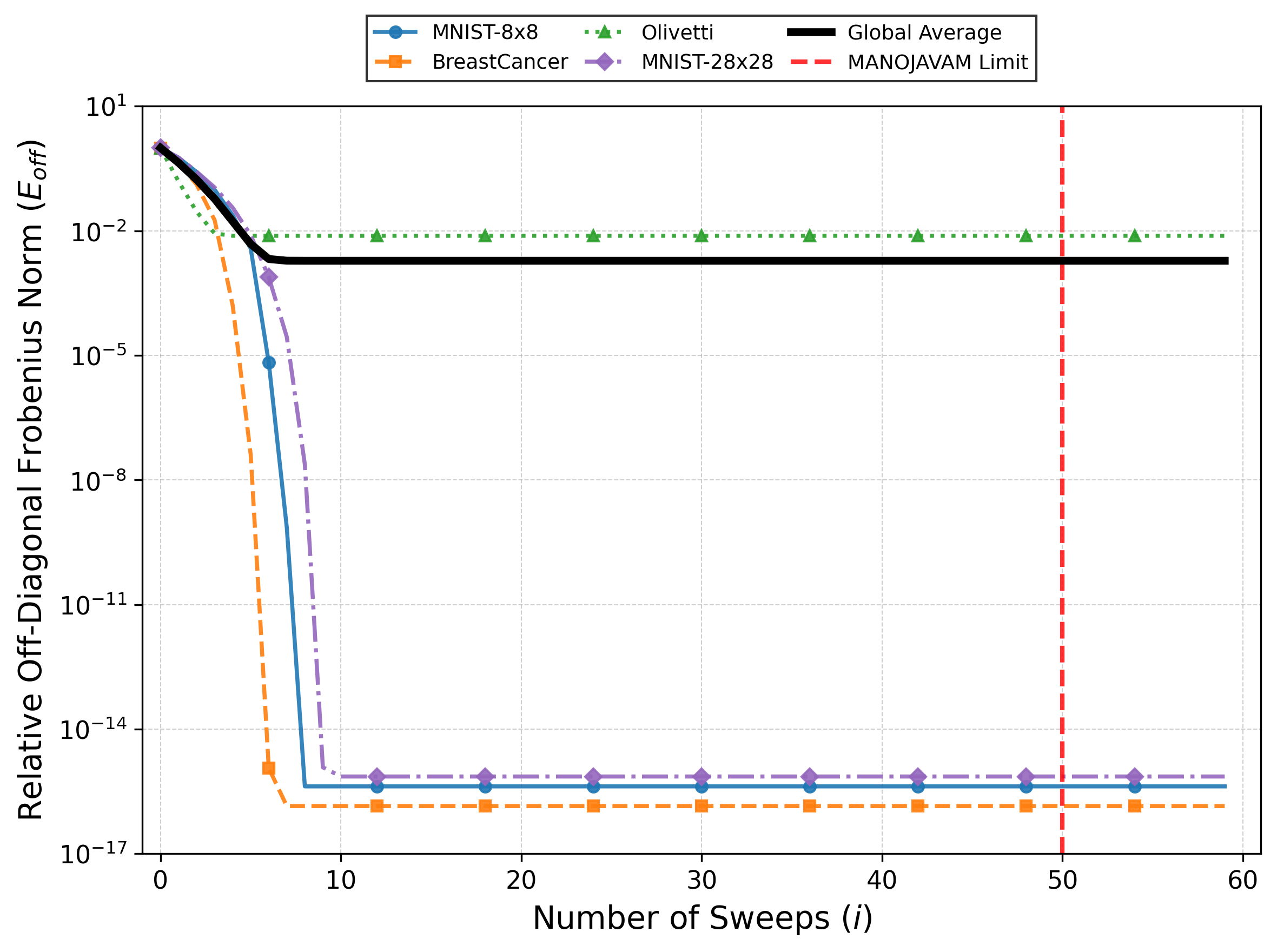}
    \caption{Relative error ($E_{off}$) vs. number of sweeps for multiple datasets}
    \label{fig:jacobi_conv_frobenius}
\end{figure}

To validate the choice of the number of Jacobi sweeps, we analyzed the relative Frobenius norm convergence across diverse datasets: MNIST, Olivetti Faces, and Breast Cancer. As illustrated in Fig.\ref{fig:jacobi_conv_frobenius}, the relative off-diagonal energy for most standard datasets saturates at the numerical noise floor within 10 to 15 iterations. 

Despite this rapid convergence in typical cases, \textit{MANOJAVAM} maintains a fixed limit of 50 iterations as a safety factor. This high upper ceiling is chosen to handle ill-conditioned datasets—where eigenvalues are closely clustered and require more rotations to achieve separation. This design choice ensures that the accelerator provides a universal "Factor of Safety," guaranteeing variance accuracy even for ill-conditioned input data distributions. Consequently, the architecture trades a negligible amount of redundant computation for a significant gain in hardware simplicity and cross-dataset reliability.
\section{Design Space Exploration}
\textit{MANOJAVAM}'s two tunable parameters include the Tile Size ($T$) and the Parallelism Index ($S$). The selection of the two parameters has a drastic impact on the performance of the accelerator. While it is favorable to choose a very large value of $S$ and $T$ to extract the highest possible execution speed, it has negative implications on resource consumption, power dissipation, and size. Thus, it is crucial to choose $S$ and $T$ based on the application and the size of the input dataset. \textit{MANOJAVAM}, being scalable, can be tailored to meet any PCA workload given its tunable tile size and cores deployed, making it extremely versatile.

The following subsections describe the variation in performance of the proposed accelerator on execution time, power consumption and resource utilization across different tile sizes and cores deployed. These ablation studies are performed on the Xilinx Virtex Ultrascale+ xcu250-figd2104-2L-e FPGA.

\subsection{Execution Time Analysis for Varying Tile Sizes}
The execution time of the accelerator varies inversely with the square of the tile size $(T^{2})$. This is due to the fact that the input dataset of dimensions $(M,N)$ is split into tiles, resulting in $\frac{MN}{T^{2}}$ tiles of computation. This is depicted in Fig.8(a), where the execution time is profiled for varying tile sizes across benchmark datasets for a fixed value of $S=32$.

\begin{figure*}[ht]
    \centering
    \subfloat[]{
        \includegraphics[width=0.47\linewidth]{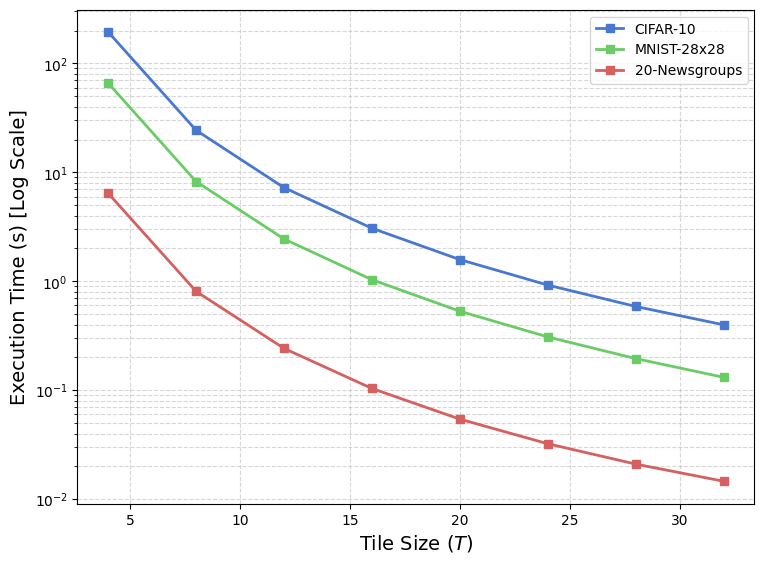}
    }
    \hfill
    \subfloat[]{
        \includegraphics[width=0.47\linewidth]{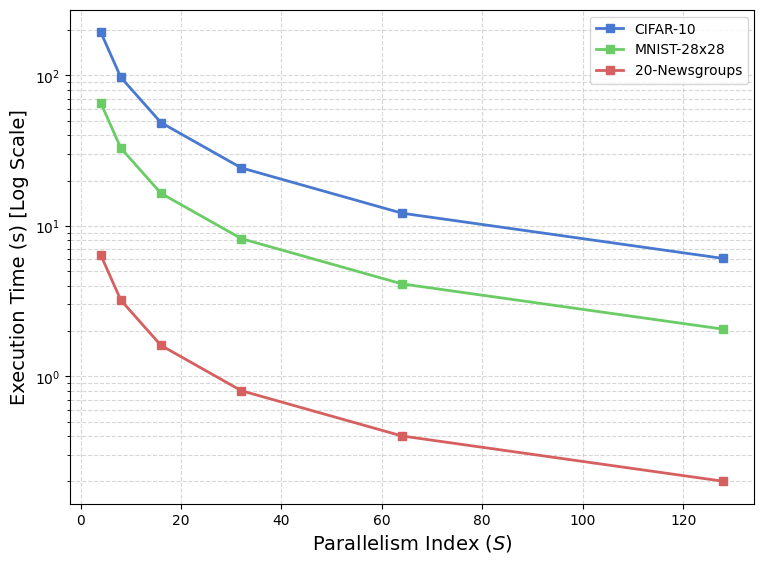}
    }
    \caption{Design Space Exploration of Architectural Latency: (a) Impact of Tile Size ($T$) on total execution time at a fixed parallelism of $S=4$; (b) Impact of Parallelism ($S$) on total execution time at a fixed tile size of $T=4$.}
    \label{fig:exectime-tilesize-parallelism}
\end{figure*}

\subsection{Execution Time Analysis for Varying Parallelism Index}
The execution time of the accelerator varies inversely with the parallelism index $(S)$. This is due to the fact that the input dataset of dimensions $(M,N)$ is split into tiles, resulting in $\frac{MN}{T^{2}}$ tiles of computation. However, as $S$ such tiles can be processed at once simultaneously, it would result in $\frac{MN}{ST^{2}}$ batches of computation. This is depicted in Fig.8(b), where the execution time is profiled for varying tile sizes across benchmark datasets for a fixed value of $T=32$.

\subsection{Power Analysis}
The power dissipated by the accelerator is different for varied values of parallelism index, $S$, and tile size, $T$. For the design space exploration on power consumption, the accelerator configuration synthesized is floorplanned. The floorplanning is such that the LHS components are assigned to one block, and each of the RHS components are assigned their separate blocks. This floorplanning ensures that the design is neatly partitioned.

Fig.9(a) illustrates the power breakdown of \textit{MANOJAVAM} as the systolic array tile size $T$ is increased from 4 to 20, while keeping the degree of parallelism $S$ fixed at 4. A clear upward trend is observed across all power components—clock, signal, logic, BRAM, and DSP—with signal power exhibiting the steepest growth, followed by logic and DSP power. This trend is architecturally driven by the quadratic increase in the number of multiply-accumulate (MAC) units per systolic array, which grows as $T^2$. As the tile size increases, the systolic array feeds in more routing interconnects into it, leading to higher switching activity and capacitive loads, especially along longer signal routes. This results in proportional increase in power dissipation. Logic power increases due to the expansion of operator feeding logic and controller FSMs. Concurrently, DSP power scales due to instantiation and operation of more MAC units. Interestingly, BRAM power consumption shows a marginal increase as the on chip memory footprint per array remains modest, with no additional banks instantiated as T scales. Overall, this analysis highlights the energy implications of scaling systolic array granularity, with larger tiles offering computational benefits at the cost of power and thermal overhead.

\begin{figure*}[!ht]
    \centering
    \hfill
    \subfloat[]{
        \includegraphics[width=0.47\linewidth]{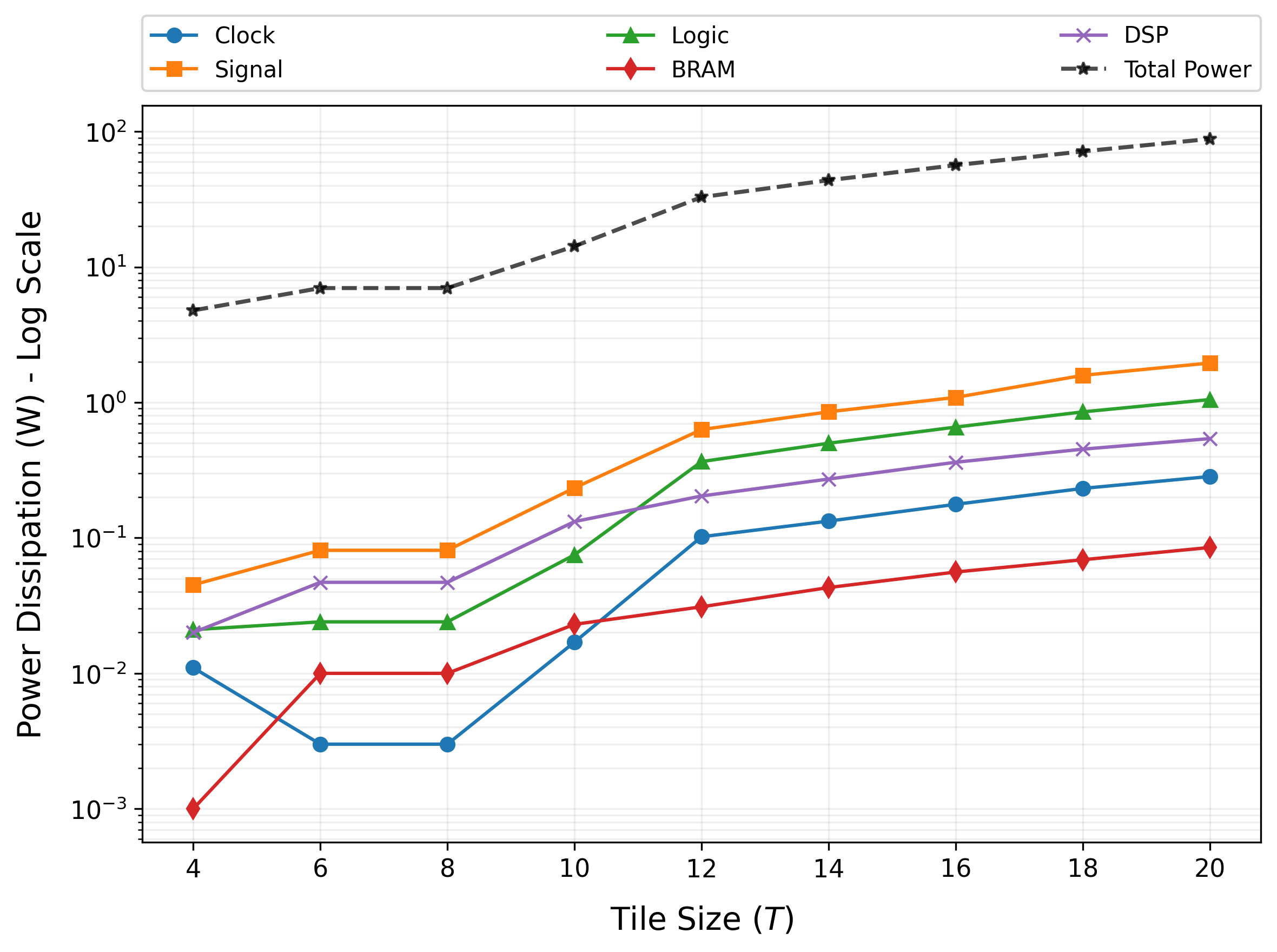}
    }
    \hfill
    \subfloat[]{
        \includegraphics[width=0.47\linewidth]{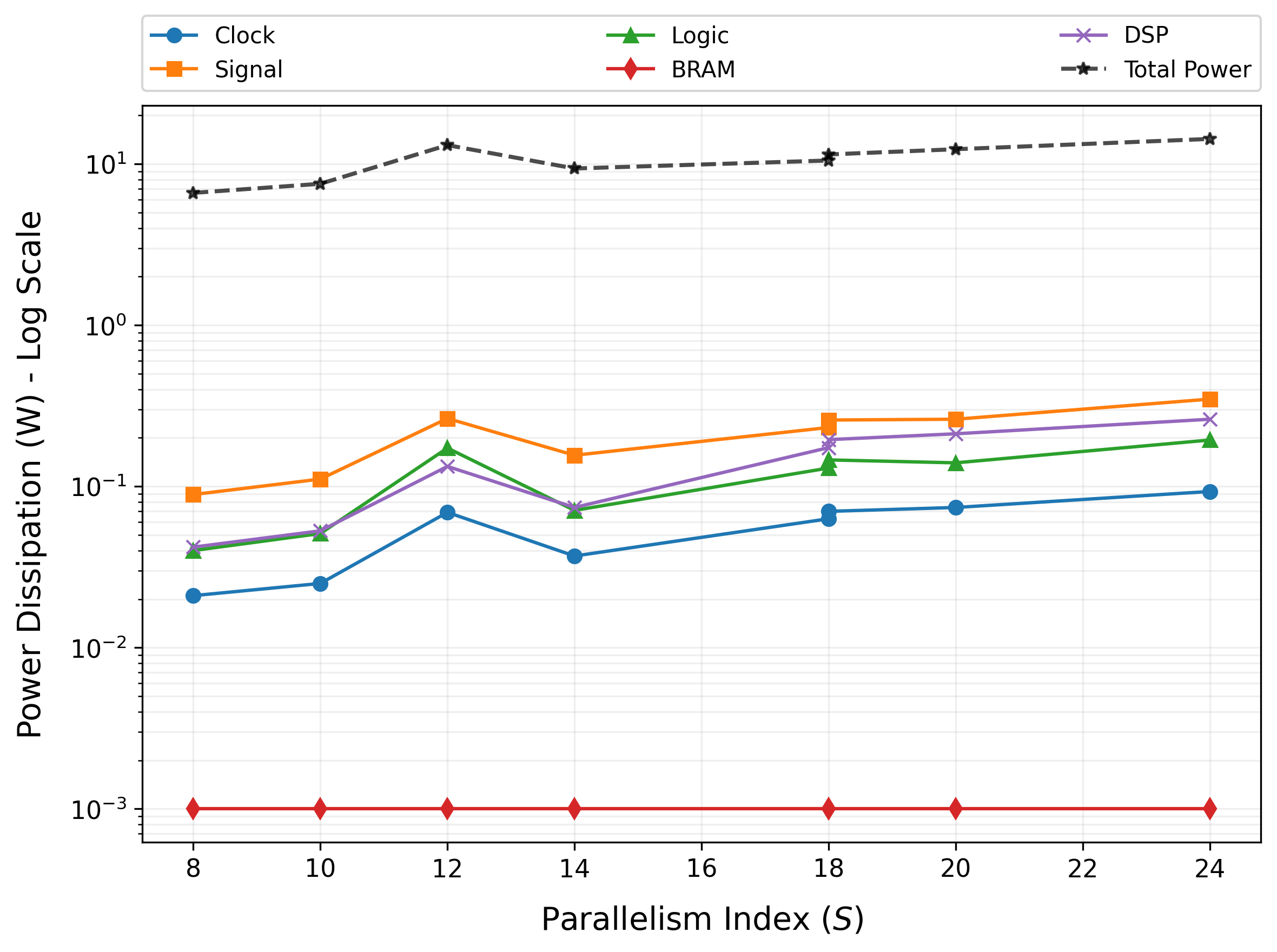}
    }
    \caption{Power Dissipation Scaling Analysis: (a) Sensitivity of power consumption to Tile Size ($T$) at a fixed parallelism of $S=4$; (b) Impact of Parallelism ($S$) on power dissipation at a fixed tile size of $T=4$}
    \label{fig:pca_exec_time_subfigs}
\end{figure*}

Fig.9(b) presents the impact of scaling parallelism index $S$ on power consumption, keeping tile size constant at $T=4$. As $S$ increases from 8 to 24, a steady rise is observed across all power components, especially signal, DSP and logic power. This trend is expected, as an increment in $S$ would instantiate a new $T$x$T$ systolic array, $T$x$T$ matrix accumulator, a private RHS cache unit and other interfacial components. DSP power increases due to the linear addition of MAC units. Signal power grows substantially due to the overhead of simultaneously feeding operands into more parallel datapaths, requiring wider operand buses and more frequent switching. Logic power rises as well, reflecting the growth in per-array control logic and operand routing infrastructure. Clock power shows a modest linear increase, which is consistent with the growing number of flip-flops and MAC units receiving the global clock. The BRAM power remains almost constant across the range of $S$ due to fixed tile size. The irregularities observed at $S=12$ and $S=14$ is due to increased routing congestion and synthesis tool optimizations respectively. Thus, the analysis in  Fig.9(b) reveals that the power dissipation increases with increase in $S$.

\subsection{Resource Consumption}
Fig.10(a) presents the FPGA resource usage profile of the accelerator, as the tile size $T$ is increased from 4 to 20 at $S=4$. All key resources - LUTs, Flip-Flops (FFs), BRAMs, and DSPs, show a monotonic increase with growing tile size. This growth is attributed to the quadratic scaling of computational units: each $T$x$T$ systolic array contains $T^{2}$ MACs, leading to proportionate increase in DSP units. This scaling also drives the increase in LUTs and FFs, required for operand buffering, pipeline registers and routing logic. The dip at $T=6$ is associated with the synthesis tool optimization logic. BRAM utilization increases steadily, as larger tiles require deeper memory for operand storage in the caches. 

\begin{figure*}[!ht]
    \centering
    \subfloat[]{
        \includegraphics[width=0.485\linewidth]{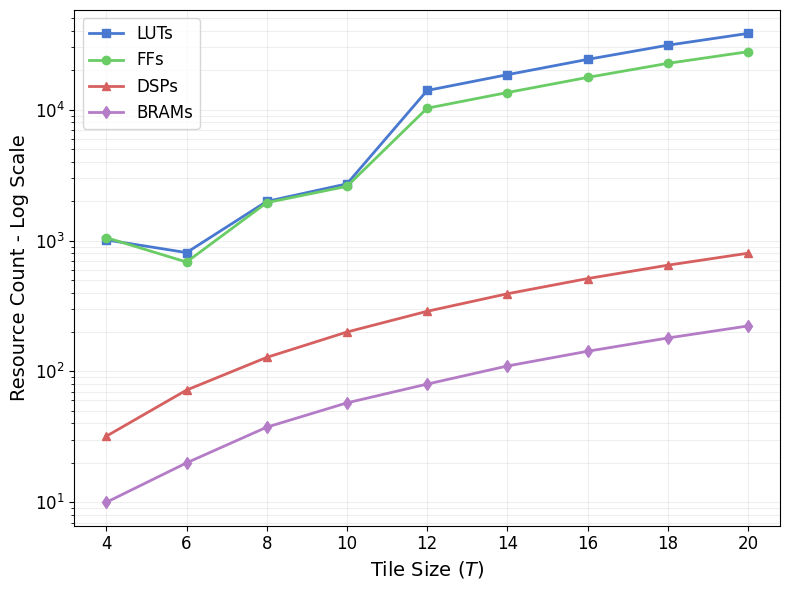}
    }
    \hfill
    \subfloat[]{
        \includegraphics[width=0.485\linewidth]{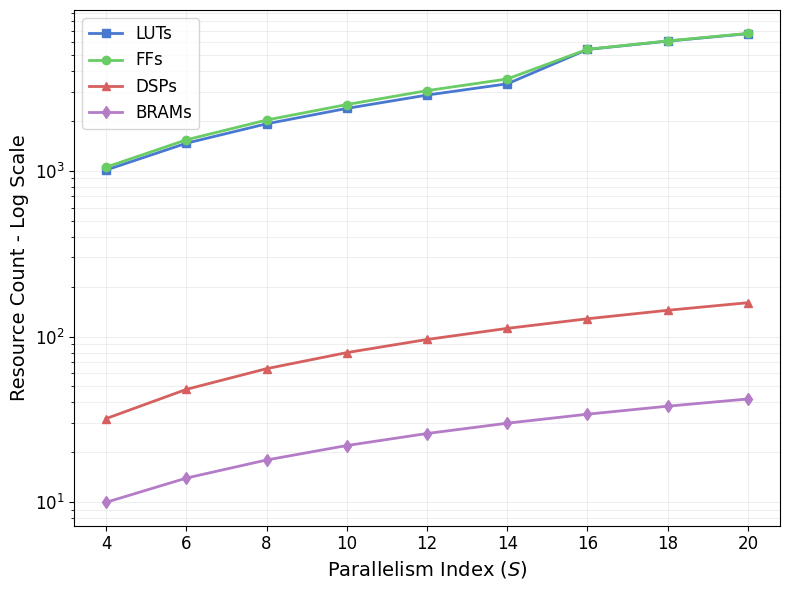}
    }
    \caption{Hardware Resource Utilization Scaling: (a) Analysis of FPGA resource requirements (LUTs/DSPs) relative to Tile Size ($T$) at $S=4$; (b) Impact of Parallelism ($S$) on resource consumption at a fixed tile size of $T=4$.}
    \label{fig:pca_exec_time_subfigs}
\end{figure*}

Fig.10(b) reflects the resource scaling of the accelerator with growing parallelism at $T=4$. The resources grow linearly due to incremental instantiation of 4x4 systolic arrays, caches, accumulators and required control logic with an increase in parallelism. The DSP usage scales predictability, as 16 MAC units get instantiated with incremental rise in $S$. LUT and Flip-Flop usage exhibit near identical scaling trends. BRAMs also increase gradually, with each array associated to private caches with operand B. Unlike tile size scaling, this form of spatial parallelism introduces isolated compute islands, resulting in more modular resource growth with minimal synthesis irregularities.
\section{Conclusion and Future Scope of Work}
This paper presented \textit{MANOJAVAM}, a domain-specific architecture tailored to accelerate the computationally intensive tasks of matrix multiplication and singular value decomposition (SVD) in Principal Component Analysis (PCA). The architecture introduces a novel block-streaming approach combined with a systolic array-based matrix multiplication engine and a highly pipelined Jacobian unit for executing the Jacobi algorithm efficiently. The dual-phase memory hierarchy, featuring mode-aware cache policies, further enhances data throughput and minimizes latency. \textit{MANOJAVAM} has been successfully realized on both FPGA and ASIC platforms, showcasing significant speedups and power savings compared to leading CPU and GPU implementations.

There are several promising directions to extending this work. \textit{MANOJAVAM} can be integrated with downstream machine learning accelerators to realize a complete on-chip analytical engine for edge-AI platforms. The current architecture can also be extended to perform adaptive covariance matrix computation to support incremental PCA. Finally, \textit{MANOJAVAM} can be deployed on real life applications, such as autonomous vehicles, bio-informatics and surveillance systems to yield insights on the edge. 

\bibliographystyle{IEEEtran}
\bibliography{manojavam}


\begin{IEEEbiography}
[{\includegraphics[width=1in,height=1.25in,clip,keepaspectratio]{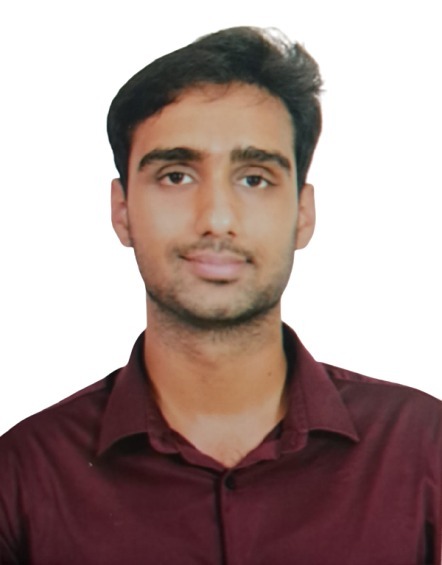}}]{Srivaths Ramasubramanian}
(Student Member, IEEE) received the B.E. degree in Electronics and Communication Engineering from Rashtreeya Vidyalaya College of Engineering (RVCE), Bengaluru, India, in 2025. He is currently pursuing the Ph.D. degree in Computer Engineering at the Donald Bren School of Information and Computer Sciences, University of California, Irvine. His research interests include hardware acceleration for machine learning and communication systems, computer architecture and FPGA-based system design.
\end{IEEEbiography}

\begin{IEEEbiography}
[{\includegraphics[width=1in,height=1.25in,clip,keepaspectratio]{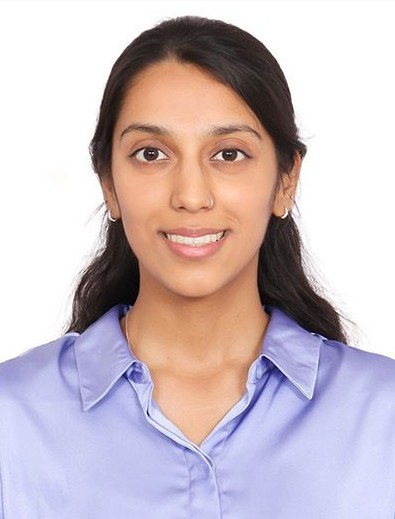}}]{Anjali Devarajan}
(Student Member, IEEE) received the B.E. degree in Electronics and Communication Engineering from Rashtreeya Vidyalaya College of Engineering (RVCE), Bengaluru, India, in 2025. She is currently an engineer at Infineon Technologies. Her research interests are embedded systems, VLSI design, and hardware acceleration for machine learning. Her experience includes internships in ADAS, control systems and low-power design. Her work has been published in IEEE and Springer conferences. 
\end{IEEEbiography}

\begin{IEEEbiography}
[{\includegraphics[width=1in,height=1.25in,clip,keepaspectratio]{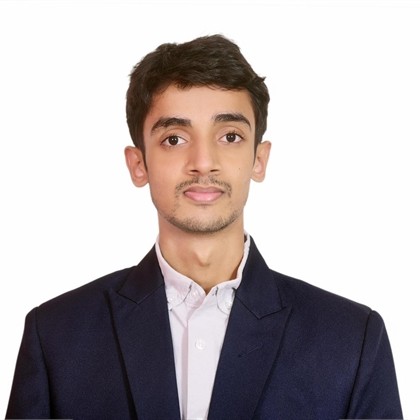}}]{Kousthub P Kaivar}
(Student Member, IEEE) received the B.E. degree in Electronics and Communication Engineering from Rashtreeya Vidyalaya College of Engineering (RVCE), Bengaluru, India, in 2025. He is currently a Hardware Engineer in the Battery Management Unit team at Enphase Energy. His interests include VLSI, Computer Architecture,  FPGA based system design, communication systems and machine learning. In addition, he has work published in communication systems involving antennas and radars in IEEE conferences. 
\end{IEEEbiography}

\begin{IEEEbiography}
[{\includegraphics[width=1in,height=1.25in,clip,keepaspectratio]{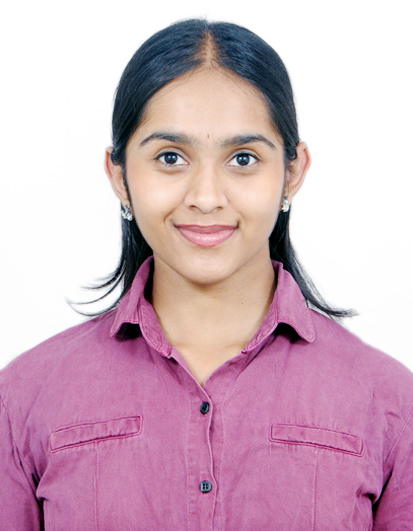}}]{Vibha Shrestta}
(Student Member, IEEE) received the B.E. degree in Electronics and Communication Engineering from Rashtreeya Vidyalaya College of Engineering (RVCE), Bengaluru, India, in 2025. She is currently working as a Design engineer at ABB. Previously, she has also worked in the research and development wing at DRDO, solving critical challenges in embedded system design and flight avionics. Her interests include VLSI, FPGA based system design, analog electronics and power electronics. 
\end{IEEEbiography}

\begin{IEEEbiography}
[{\includegraphics[width=1in,height=1.25in,clip,keepaspectratio]{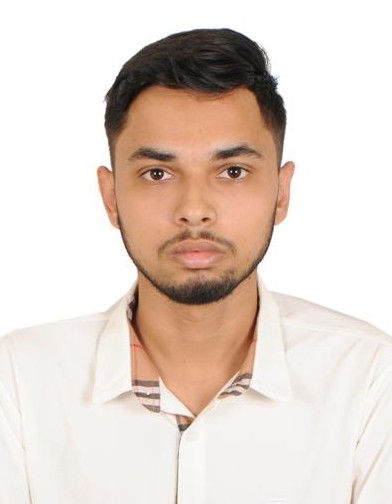}}]{Shashank D}
(Student Member, IEEE) received the B.E. degree in Electronics and Communication Engineering from Rashtreeya Vidyalaya College of Engineering (RVCE), Bengaluru, India, in 2025. He is currently working as an Associate Sales Development Representative at Astar Data LLP . His interests include VLSI,  FPGA based system design, communication systems and Machine Learning.
\end{IEEEbiography}

\begin{IEEEbiography}
[{\includegraphics[width=1in,height=1.25in,clip,keepaspectratio]{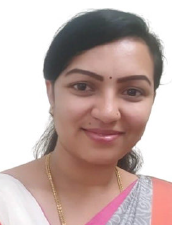}}]{Dr. Sowmyarani C.N}
(Member, IEEE) received the Ph.D. degree in data privacy. She is currently a Professor with the Department of Computer Science and Engineering, R. V. College of Engineering. She has 16 years of teaching experience. She has published more than 60 research publications. She has delivered many talks and training sessions in the field of data privacy, networks, and security. She has been working on research projects funded by VGST, Government of Karnataka, India. Her research interests include data engineering, data privacy, computer networks, cyber security and education technology.
\end{IEEEbiography}

\vspace{-5.0cm}
\begin{IEEEbiography}
[{\includegraphics[width=1in,height=1.25in,clip,keepaspectratio]{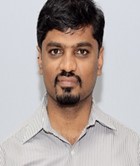}}]{Dr. Govinda Raju M}
received the B.E. degree in Electronics and Communication Engineering from Visveswaraya Technological University(VTU), Karnataka, India, in 2004, and the M.Tech. and Ph.D. degrees in Electronics from VTU in 2007 and 2013, respectively. He is currently an associate professor in the Department of Electronics and Communication Engineering at RV College of Engineering, Bengaluru, India. His research interests include embedded system design, computer architecture, real-time systems, and embedded automotive systems. He has published over 50  research papers in reputed journals and conferences. He has guided several postgraduate and undergraduate student projects funded by national bodies.    
\end{IEEEbiography}   

\vspace{-5.0cm}
\begin{IEEEbiography}
[{\includegraphics[width=1in,height=1.25in,clip,keepaspectratio]{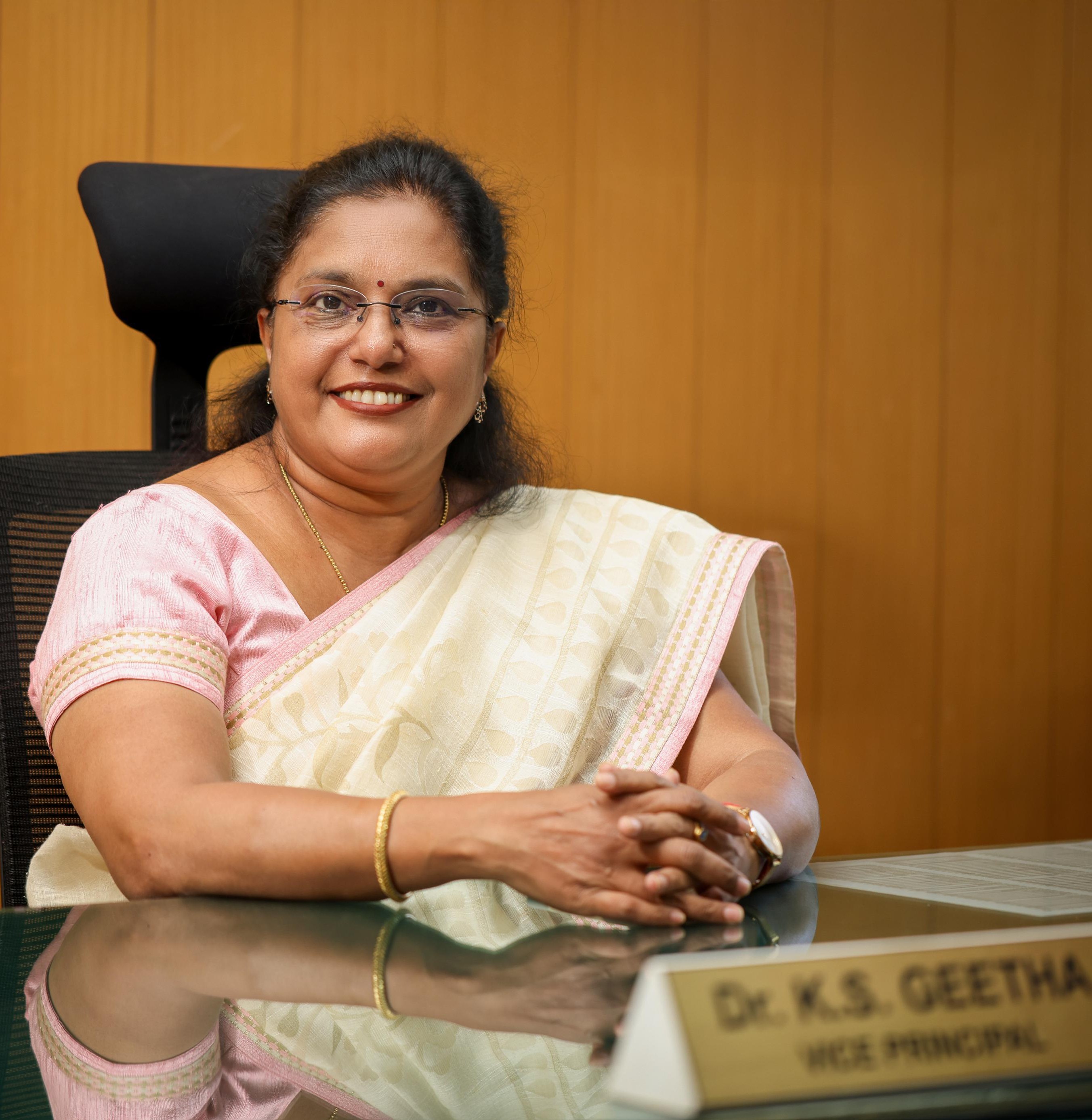}}]{Dr. K.S Geetha}
(Senior Member, IEEE) received the B.E. degree in Electronics and Communication Engineering in 1991 and the M.Tech. degree in Computer Applications to Industrial Drives in 1998 from the National Institute of Engineering, Mysore, India, and the Ph.D. degree from Visvesvaraya Technological University, India, in 2012. She is a professor in the department of Electronics \& Communication Engineering and currently  the Vice Principal of Rashtreeya Vidyalaya College of Engineering (RVCE), Bengaluru, India. Her research interests include image processing, signal processing, VLSI design, and flexible electronics.
\end{IEEEbiography}

\end{document}